\documentclass[preprint,showpacs,preprintnumbers,amsmath,amssymb]{revtex4}

\usepackage{graphicx}
\usepackage{dcolumn}
\usepackage{bm}

\begin{document}

\title{Space-and Time-like Electromagnetic Kaon Form Factors}

\author{Udit Raha}%
 \email{udit@phys.ntu.edu.tw}
 \affiliation{%
Department of Physics and Center for Theoretical Sciences, 
National Taiwan University, Taipei 10617, Taiwan
}%
\author{Hiroaki Kohyama}%
 \email{kohyama@phys.sinica.edu.tw}
 \affiliation{%
Institute of Physics, Academia Sinica, Taipei, Taiwan 115, Republic of China
}
\affiliation{%
Physics Division, National Center for Theoretical Sciences,
Hsinchu, Taiwan 300, Republic of China
}%

\date{\today}


\begin{abstract}
\noindent A simultaneous investigation of the space- and time-like
electromagnetic form factors of the charged kaon is presented within the 
framework of light-cone QCD, with perturbative $k_T$-factorization 
including Sudakov suppression. The effects of power suppressed sub-leading 
twists and the genuine ``soft'' QCD corrections turn out to be dominant at 
low- and moderate-energies/momentum transfers $Q^2$. Our predictions agree
well with the available moderate-energy experimental data, including the
recent results from the CLEO measurements and certain estimates based on 
the phenomenological analyses of $J/\psi$ decays.    
\end{abstract}
 

\maketitle


\section{Introduction}
Electromagnetic form factors are interesting physical observables in hadronic physics
which directly provide insights into the hadronic constituents, charge distributions, 
currents, color and flavor within the hadrons. Their precise knowledge is of fundamental 
importance for a realistic and accurate description of exclusive nuclear reactions
that serve as ideal testing grounds for understanding the dynamics of confinement in 
QCD that have been grappling with physicists ever since the discovery of asymptotic freedom. 

In the last few decades, there has been significant experimental efforts in extracting hadronic 
form factors  (e.g., see. \cite{pion_exp,ee-hh,CLEO} for the charged pion form factors ) 
from various exclusive processes. However, in case of the charged kaon form factors their 
behavior were very severely constrained due to absence of reliable experimental data. 
Since the mid-90s kaon photo-/electro-production experiments on reactions such as 
$A(\gamma,K)YB$ and $A(e,e'K)YB$  (target $A$, produced hyperon $Y$
and recoil $B$) have invited  some renewed interest in the study of
kaon form factors, although the existing data is still too limited, restricted only
to the very low space-like region, as low as $-q^2\leq 0.2$ GeV$^2$ \cite{kaon_exp_space}. 
In the time-like region, there are more scattered data up to several GeV$^2$ 
(albeit with very large error-bars) for time-like processes such as $\gamma^* \rightarrow K^+ K^-$, 
extracted from annihilation reactions such $J/\psi\rightarrow e^+e^-\rightarrow h^+ h^-$  $(h=\pi,K,...)$
by applying suitable experimental cuts. A compilation of previously extracted kaon form factors 
for $q^2=Q^2<10$ GeV$^2$ is given in \cite{ee-hh}. Recently, high precision
measurements by the CLEO collaboration with first ever identified time-like kaons for 
$Q^2>4$ GeV$^2$ have yielded the following results: $\left\vert G_K (13.48\,{\rm GeV}^2)\right\vert=0.063\pm 
0.004$(stat)$\pm 0.001$(syst) and $Q^2\left\vert G_K (13.48\,{\rm GeV}^2)\right\vert=0.85\pm 0.05$(stat) 
$\pm 0.02$(syst) GeV$^2$ \cite{CLEO}. Note that in this paper, we shall use the symbol $G_K$ 
for the time-like kaon form factor to distinguish it from the space-like counterpart $F_K$.

Notwithstanding the aforementioned problem of paucity of quality statistics 
of the existing kaon data, the purpose of the paper is an effort to make a 
prediction for the charged kaon form factors using the framework of
perturbative factorization \cite{Brodsky,Radyushkin}. In this way, 
we hope to throw some light on their possible behavior, especially, 
at the phenomenological intermediate energy region, where significant 
non-perturbative effects tend to spoil the asymptotic perturbative QCD 
(pQCD)  results like the celebrated quark counting rule
that predicts the scaling behavior $\{F,G\}_K(Q^2)\sim 1/Q^2$ \cite{Brodsky,FJ}. 
Analyses of the pion form factors convincingly show that the standard pQCD
with only twist-2 effects are much too small to explain the currently
available experimental data at low- and moderate-energies. This calls for the
inclusion of non-perturbative corrections from the genuine ``soft'' QCD 
\cite{Nest,Isgur,Stefanis1,Stefanis2,Bakulev} and the sub-leading twists 
that can give rise to unnaturally large contributions at moderate range of
$Q^2$-values. in particular, twist-3 enhancements were seen to be quite large
in the previous studies for the space-like pion form factor 
\cite{GT,Pasu,Wei,Huang,Raha1}, the space-like kaon form factor
\cite{Raha1,Wu}, and in the studies of $B\rightarrow\pi$ transition form 
factors \cite{Keum,Sanda,Li,Lu}. In fact, the analysis presented in \cite{Raha2}
shows a possible scenario where the contributions from the twist-3 terms in 
the time-like region can turn out to be exceptionally large.  This seemed to
resolve the bulk of the existing  experimental discrepancy between the space-
and the time-like pion data.

In this paper, following \cite{Raha1,Raha2} we extend the analysis to the
space- and the time-like kaon form factors, where in addition to the twist-2
and twist-3 terms we explicitely include twist-4 corrections. Thereby, we
show that the large twist-3 contributions are indeed a non-trivial aspect 
of our result in comparison with the other twist contributions, e.g., the 
2-particle twist-4 contributions are explicitly shown to be about a third
of the magnitude of the twist-2 terms. The paper is organized as follows: the 
second section briefly reviews the theoretical background, the third 
section deals with the details of our numerical analysis, results and
discussions of the essential features of our results, and finally, we give 
our conclusions. For the purpose of book keeping, we provide a collection 
of relevant mathematical formulas in the appendix.

\section{Hard and Soft Kaon Form Factors}
\subsection{Factorized pQCD}
The basic definitions of the space- and time-like electromagnetic form factors
are given in terms of the following local matrix elements of the electromagnetic 
quark currents $J^{\rm em}_\mu$:
\begin{eqnarray}
e(P '+P)_\mu\, F_K(Q^2)\!\!&=&\!\!\langle K^{\pm}(P ')|J^{\rm em}_\mu(0)|K^{\pm}(P)\rangle\,;\nonumber\\
e(P '-P)_\mu\, G_K(Q^2)\!\!&=&\!\!\langle K^{+}(P ')\,K^{-}(P)|J^{\rm em}_\mu(0)|0\rangle\,;\nonumber\\
J^{\rm em}_\mu\!\!&=&\!\!\sum^{\,}_{f}e_f\bar{q}_f\gamma_\mu q_f\,,
\end{eqnarray}
where $e$ is the electronic charge and $f$ is the flavor of the valence quark 
$q_f$ with charge $e_f$. In terms of light-cone co-ordinates, 
$P=(Q/\sqrt{2},0,{\mathbf 0}_{T})$ and $P'=(0,Q/\sqrt{2},{\mathbf  0}_{T})$ 
are the incoming and outgoing external kaon momenta in the Breit-frame. 
In the space-like domain, $q^2=(P'-P)^2=-Q^2\leq 0$, whereas for the time-like 
domain $q^2=(P'+P)^2=Q^2\geq 0$. Here, $Q$ is assumed to be 
much larger compared to the kaon mass $m_K$, so that $P$ and $P'$ almost lie along the 
light-cone directions.

\begin{figure}
    \begin{tabular}{cc}
      \hspace{0cm}\resizebox{15cm}{!}{\includegraphics{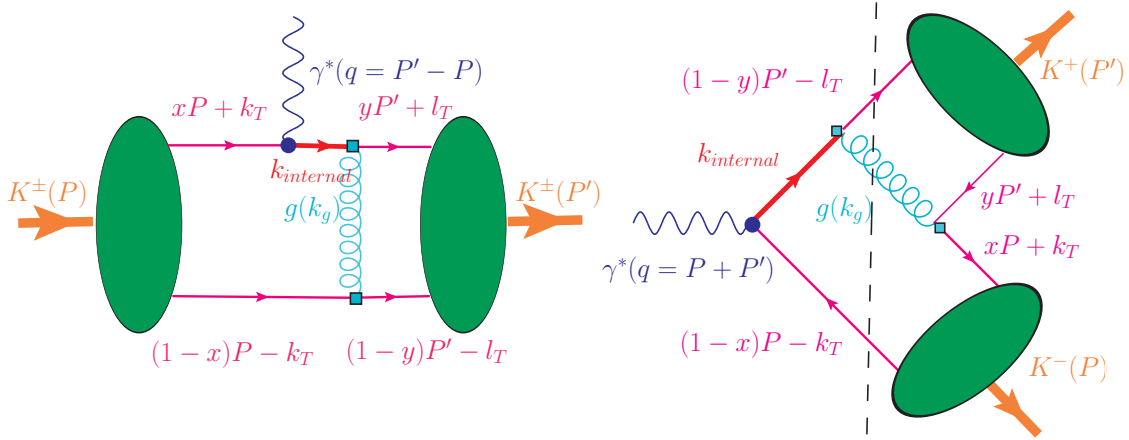}}
    \end{tabular}
    \caption{LO Feynman diagrams in pQCD for hard contributions to the
      charged kaon form factors in the space-like (left) and the time-like (right)
      region.}
    \label{fig:feyndiag}
\end{figure}

In our approach, the total contributions to the charged kaon form factors 
come from the factorizable "hard" parts $\{F,G\}^{\rm hard}_K(Q^2)$ calculable  
within a perturbative framework, and the non-factorizable soft parts 
$\{F,G\}^{\rm soft}_K(Q^2)$ relying on some non-perturbative techniques. 
The calculation of the hard parts rest on the essential assumption that 
at suitable high energy scales, the form factors are {\it factorizable}, i.e., 
separable into parts dominated by short- and long-distance dynamics. The 
short-distance dynamics are represented by the kernel of interactions between 
highly off-shell partons, above the so-called {\it factorization scale}
$\mu_F$. While, the long-distance dynamics below the factorization scale are 
implicitly encoded within the kaonic wavefunctions/distribution amplitudes
(DAs) with near-on-shell partons. Note that due to the well-known {\it impulse} 
or {\it frozen} approximation applicable for all high-energy exclusive mechanisms, 
the dominant contributions come entirely from the leading order (LO)  Fock state, i.e., 
a $q\bar{q}$ valence quark configuration. The higher Fock states are neglected
with contributions relatively suppressed by higher powers of $1/Q^2$. 
Fig.~\ref{fig:feyndiag} shows two representative Feynman diagrams 
(there are 4 diagrams each for the space-and time-like cases) with LO hard
kernels each having a single hard gluon exchange. These are convoluted with 
the incoming and outgoing kaon DAs to obtain the hard factorized kaon form 
factors. In this analysis, we calculate $\{F,G\}^{\rm hard}_K(Q^2)$ up to 
twist-4 accuracy in the two-particle sector, including explicit transverse 
momentum ($k_T$) dependence (TMD) of the constituent valence partons. 
The non-factorizable soft contributions, on the other hand, can either be 
calculated using Drell-Yan-West type of wavefunctions overlap ansatz
\cite{DYW}, or from QCD sum rules (QCDSR) incorporating local quark-hadron 
duality (LD) principle \cite{Nest}. Both these approaches to parameterize the 
genuine soft contributions are known to give very similar results. In this 
work, we follow  the latter approach. The above assertion can then be 
summarized by
\begin{eqnarray}
\{F,G\}_K(Q^2)\!\!&=&\!\!\{F,G\}^{\rm soft}_K(Q^2)+\{F,G\}^{\rm
  hard}_K(Q^2)\,; \nonumber \\
\{F,G\}^{\rm hard}_K(Q^2)\!\!&=&\!\!\Delta\{F,G\}^{\rm twist2}_K(Q^2)+\Delta\{F,G\}^{\rm twist3}_K(Q^2)+\Delta\{F,G\}^{\rm twist4}_K(Q^2)\,. 
\end{eqnarray}
The principal  inputs for determining the factorized kaon form factors are the 
collinear/light-cone DAs which encode all the non-perturbative physics. They 
are universal in nature (frame or process independent), in a sense, once 
they are determined at a certain process, they could yield predictions for another. 
To next-to-leading order (NLO) in conformal twist expansion there are one 2-particle 
twist-2 DA $\phi_{2;K}$ with an axial-vector structure, two 2-particle twist-3 
DAs - $\phi^p_{3;K}$ with a pseudo-scalar structure and $\phi^\sigma_{3;K}$ 
with a pseudo-tensor structure, and finally, two 2-particle twist-4 DAs  
${\mathbb{A}}_{4;K}$ and  ${\mathbb{B}}_{4;K}=g_{4;K}-\phi_{2;K}$ both having
pseudo-scalar structures \cite{Filyanov,Ball,Lenz}. As an example for $K^-$,
we display the twist-2 DA in terms of the following pseudo-scalar matrix
element with $\xi=2x-1$: 
\begin{eqnarray}
\langle0\left|\bar{u}(z)\,\gamma_\mu \gamma_5\,
s(-z)\right|K^-(P)\rangle=iP_\mu\,\int^1_0 dx\,
e^{i\xi(P z)}\,\phi_{2;K}(x,\mu^2)\,,
\end{eqnarray}
with the normalization condition
\begin{equation}
N_{2;K}=\int^1_0\,\phi_{2;K}(x,\mu^2)\, dx = \frac{f_K}{2\sqrt{2N_c}}\,,
\end{equation}
where $f_K$ is the kaon decay constant defined in the local limit $z\rightarrow0$ by
\begin{equation}
\left<0\left|\bar{u}(0)\,\gamma_\mu \gamma_5 \,s(0)\right|K^-(P)\right> =if_K
P_\mu\,.
\end{equation}
In the above equations, $x$ is the collinear/light-cone momentum fraction 
($x_i=k^+/P^+$) carried by the individual valence quarks ($x$ for the $s$ 
quark and $\bar{x}=1-x$ for the anti-quark $\bar{u}$.)  Note that the 
gauge-connection factor in the above matrix element is assumed implicitly. 
To the leading logarithmic accuracy $\phi_{2;K}$ satisfies the well-known 
ER-BL evolution equation \cite{Brodsky,Radyushkin} and can be expressed as 
an irreducible representation of the special collinear conformal group 
S${\mathbb{L}}(2,{\mathbb{R}})$, in terms of standard {\it Gagenbauer 
Polynomials} $C^{3/2}_n(\xi)$:
\begin{equation}
\label{eq:DA_twist2}
 \phi_{2;K}(x,\mu^2)=\phi^{(\rm as)}_{2;K}(x)\,\sum^{\infty}_{n=0} a^K_n(\mu^2_0)
  \,C^{3/2}_n(\xi)\left(\frac{\alpha_s(\mu^2)}{\alpha_s(\mu^2_0)}\right)^{-4\gamma^{(0)}_n/9}
 + {\mathcal O}(\alpha_s)\,,
\end{equation}
with the asymptotic twist-2 DA given by
\begin{eqnarray}
\label{eq:DA_twist2asy}
\phi^{(\rm as)}_{2;K}(x)=\phi_{2;K}(x,\mu^2\rightarrow\infty)=\frac{3f_{K}}{\sqrt{2N_c}}\,x(1-x)\,.
\end{eqnarray}
The standard QCD $\overline{\rm{MS}}$ running coupling $\alpha_s(\mu^2)$ to
two-loop accuracy is given by
\begin{equation}
\label{eq:coupling}
\frac{\alpha_s(\mu^2)}{\pi}=\frac{1}{\beta_0\ln(\mu^2/\Lambda^2_{\rm QCD})}-\frac{\beta_1\ln(\ln(\mu^2/\Lambda^2_{\rm
    QCD}))}{\beta^3_0\ln^2(\mu^2/\Lambda^2_{\rm QCD})}
 \end{equation}
with $\Lambda_{\rm QCD}\approx0.2$ GeV, $\beta_0\!=(11N_c-2N_f)/12=\!9/4$ and 
$\beta_1\!=(51N_c-19N_f)/24=\!4$ for $N_c\!=\!N_f\!=\!3$. The ratio of the QCD 
couplings represents the renormalization group (RG) evolution of the Gagenbauer
moments  $a^K_n$ from the normalization scale $\mu_0\approx1$ GeV to the
generic scale $\mu$, with LO anomalous dimensions given by
\begin{eqnarray}
\gamma^{(0)}_n=\frac{4}{3}\left\{\frac{1}{4}+\sum^{n+1}_{k=2}\frac{1}{k}-\frac{1}{2(n+1)(n+2)}\right\}\geq 0
\end{eqnarray}
The Gagenbauer moments represent the genuine non-perturbative inputs to the DAs and 
are usually determined using lattice simulations or from light-cone sum rules
(LCSR). In this work, we use the latter inputs, since the moments for the
higher twist DAs are yet to be determined precisely in Lattice QCD. Note that 
the lower order moments in both approaches are known to be in good agreement 
with each other. However, dealing with such an infinite number of terms in the
non-asymptotic DAs become a matter of technical challenge as the higher order 
moments are extremely difficult to determine. Hence, for practical simplicity 
of calculation, one truncates the DA series up to the first couple of terms 
only. Moreover, the increasing anomalous dimensions tend to suppress the
higher order terms. In this analysis, we consider the series up to the term 
with the second moment $a^K_2$. The rest of the non-asymptotic collinear DAs, 
i.e., the 2-particle twist-3 and twist-4 DAs which we also consider in this 
work, have more elaborate expressions and are, therefore, relegated to the 
appendix along with their RG evolutions. A summary of the relevant DA
parameters determined from LCSR at the normalization scale of $\mu_0\approx1$ 
GeV is presented in Table.~\ref{table:para}.

\begin{figure}
     \begin{center}
     \resizebox{5cm}{!}{\includegraphics{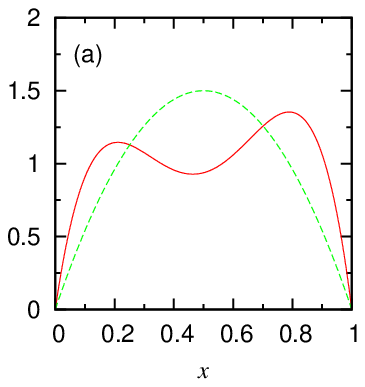}} \\
     \resizebox{5cm}{!}{\includegraphics{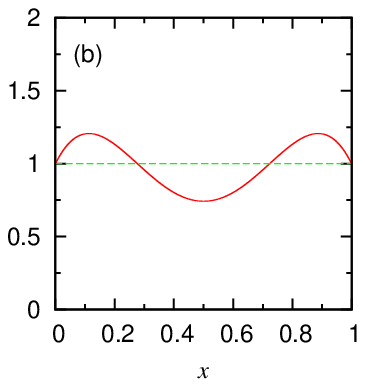}} \quad
     \resizebox{5cm}{!}{\includegraphics{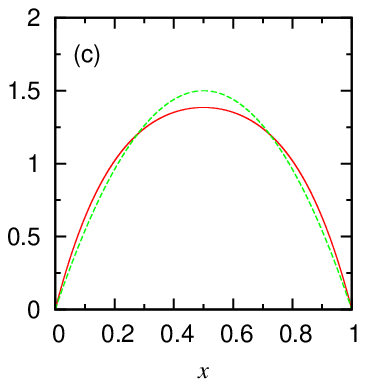}} \\
     \resizebox{5cm}{!}{\includegraphics{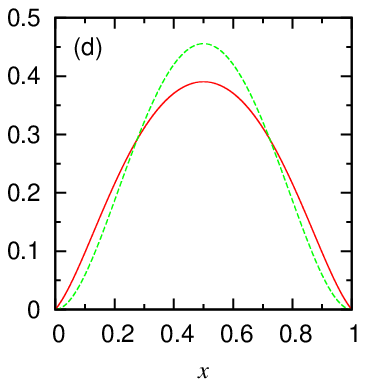}}\,\,\quad \quad
     \resizebox{4.7cm}{!}{\includegraphics{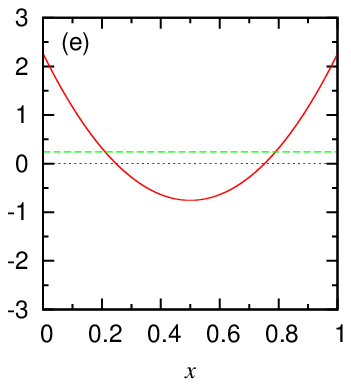}} 
     \end{center}
     \caption{The collinear twist-2 and the 2-particle twist-3 and twist-4 
      DAs for the kaon (modulo, the normalizations $N_{t;K}$), 
      shown using solid (red) lines, i.e., (a) $\phi_{2;K}(x,\mu_0^2)$, 
      (b) $\phi^{\,p}_{3;K}(x,\mu_0^2)$, (c) $\phi^{\,\sigma}_{3;K}(x,\mu_0^2)$, 
      (d) $m^2_K{\mathbb A}_{4;K}(x,\mu_0^2)$, and  (e) $m^2_Kg_{4;K}(x,\mu_0^2)$, 
      along with their respective asymptotic DAs, shown using dashed (green) 
      lines. The DAs are defined at the scale $\mu_0=1$ GeV.}
    \label{fig:KDAs}
\end{figure}

\begin{figure}
     \begin{center}
     \resizebox{5cm}{!}{\includegraphics{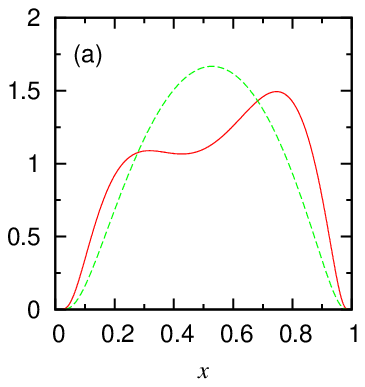}} \\
     \resizebox{5cm}{!}{\includegraphics{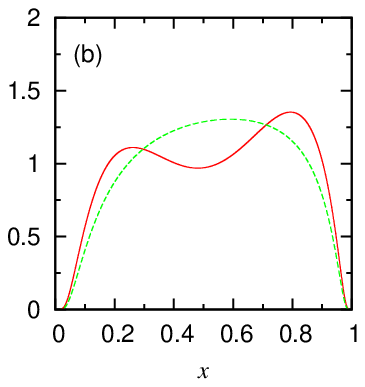}} \quad
     \resizebox{5cm}{!}{\includegraphics{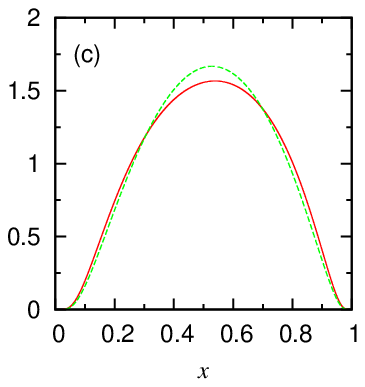}} \\
     \resizebox{5cm}{!}{\includegraphics{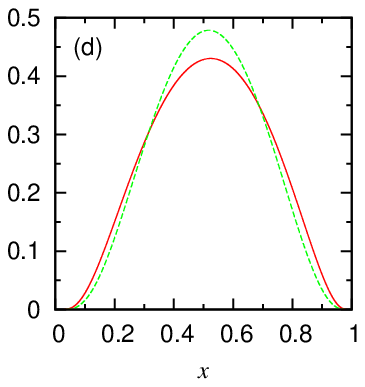}}\,\,\quad \quad
     \resizebox{4.7cm}{!}{\includegraphics{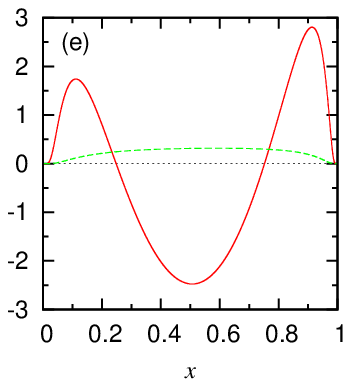}} 
     \end{center}
     \caption{The BHL modified twist-2 and the 2-particle twist-3 
      and twist-4 DAs for the kaon (modulo, the normalizations $N_{t;K}$), 
      shown using solid (red) lines, i.e., (a) $\tilde{\mathcal
        P}_{2;K}(x,\mu_0^2)$, (b) $\tilde{\mathcal P}^{\,p}_{3;K}(x,\mu_0^2)$, 
      (c) $\tilde{\mathcal P}^{\,\sigma}_{3;K}(x,\mu_0^2)$, (d)
      $m^2_K\tilde{\mathcal P}^{\mathbb A}_{4;K}(x,\mu_0^2)$, and  (e) 
      $m^2_K\tilde{\mathcal P}^g_{4;K}(x,\mu_0^2)$, along with their 
      respective asymptotic DAs, shown using dashed (green) lines. 
      The DAs are defined at the scale $\mu_0=1$ GeV.}
    \label{fig:KBHLDAs}
\end{figure}

A common feature of DAs derived from LCSR is that  they are {\it endpoint } 
dominated due to large kinematic enhancements when the light-cone momentum
fractions tend to the endpoints (i.e., $x\rightarrow 0,1$). One possible way
to suppress such artificial enhancement is to use the Brodsky-Huang-Lepage 
(BLH) Gaussian parameterization \cite{BHL}, where the intrinsic TMD of the 
valance partons within the full kaon wavefunction $\Psi_{t;K}$  (for each 
twist $t=2,3,4$) is explicitly modeled by including an addition wavefunction 
$\Sigma_{t;K}$, i.e., 
\begin{eqnarray}
\Psi_{t;K}(x,{\mathbf k}_{T},\mu^2,{\mathcal M}_{\{u,d,s\}})=A_{t;K}\,\phi_{t;K}(x,\mu^2)\,\Sigma_{t;K}(x,{\mathbf k}_{T},{\mathcal M}_{\{u,d,s\}})\,,
\end{eqnarray}
where the form of $\Sigma_{t;K}$ is chosen similar to that of a harmonic 
oscillator wavefunction that can maximally suppress such endpoint effects 
and given by
\begin{eqnarray}
\Sigma_{t;K}(x,{\mathbf k}_{T},{\mathcal M}_{\{u,d,s\}})=\frac{16\pi^2\beta_{t;K}^2}{x(1-x)}\,{\rm{exp}}\left[-\beta^2_{t;K}\left(\frac{{\mathcal M}^2_{s}+{\mathbf k}^2_{T}}{x}+\frac{{\mathcal M}^2_{u,d}+{\mathbf k}^2_{T}}{1-x}\right)\right]\,.
\end{eqnarray}
Note that the constituent quark masses ${\mathcal M}_{\{u,d,s\}}$ are 
introduced to parameterize the QCD vacuum effects, while the parameters 
$A_{t;K}$ and  $\beta_{t;K}$ for the individual twists are phenomenologically 
extracted as described in the next section (also see, \cite{Raha1}). Next to 
obtain the full TMD modified kaon DAs $\tilde{{\mathcal P}}_{t;K}$ in the impact parameter 
or $b$-representation, we use the Brodsky-Lepage definition of the DA 
\cite{Brodsky}, yielding
\begin{eqnarray}
\tilde{{\mathcal P}}_{t;K}(x,b,\mu^2,{\mathcal M}_{\{u,d,s\}})\!\!&=&\!\!\int^{1/b^2}_{0}\frac{d^2{\mathbf k}_{T}}{16\pi^3}
\,\Psi_{t;M}(x,{\mathbf k}_{T},\mu^2,{\mathcal M}_{\{u,d,s\}})\nonumber\\
&=&\!\!{A_{t;K}}\,\phi_{t;K}(x,\mu^2)\,{\rm{exp}}\left[-{\beta^2_{t;K}}\left(\frac{{{\mathcal
        M}^2_{s}}}{x}+\frac{{{\mathcal
        M}^2_{u,d}}}{1-x}\right)\right]\nonumber\\
&&\qquad\times{\rm{exp}}\left[-\frac{b^2x(1-x)}{4{\beta^2_{t;K}}}\right]\,.
\end{eqnarray}
In Figs.~\ref{fig:KDAs} and \ref{fig:KBHLDAs}, we show the various 
collinear DAs (which are endpoint enhanced) and the modified BHL 
DAs (which are endpoint suppressed) respectively. We also 
display the corresponding  asymptotic forms of the DAs.

The inclusion of the TMD in the hard scattering kernel at the same time also 
serves as a natural regulator for possible endpoint enhancements. However,
this leads to the appearance of large logarithms in the kernel due to
incomplete cancellation between soft gluon bremsstrahlung and radiative 
corrections that may spoil the perturbative convergence and, hence,
the validity of the collinear factorization. While, the large 
single-logarithms such as $\alpha_s\ln Q^2$  can be effectively tackled  
using usual RG techniques like UV divergences, the large double-logarithms 
or {\it Sudakov} logarithms involving TMD such as $\alpha_s\ln^2(Q^2/k^2_T)$, 
arising from the overlap of the leading soft and collinear kinematic regions 
of radiative gluons, can not be similarly handled in ordinary fixed order 
perturbation theory. The alternative is to use resummation techniques to 
all orders in the strong coupling constant $\alpha_s$  which organizes the 
double-logarithms within exponential Sudakov factors to get eventually 
systematically absorbed by a re-definition of the DAs. Such Sudakov factors 
represent the perturbative tail of the DAs and suppress non-perturbative 
enhancement that arise from constituent partonic configurations which involve 
large impact space separations. For a review of the Sudakov form factors and 
their application to exclusive physics, the reader is referred to 
\cite{Stefanis2,Lu,Botts,LiSterman}. There may be other radiative collinear
double-logarithms such as $\alpha_s\ln^2x$ which may be resummed using {\it threshold 
resummation} \cite{Sanda,Li} to suppress additional collinear enhancements in
the kernel. The threshold resummation along with the Sudakov resummation arising 
from different subprocesses in pQCD factorization provides natural suppression to the 
endpoint and other non-perturbative enhancements and are relevant in the range
of currently probed energy/momentum transfer values. The upshot is that the hard
perturbative contributions are enhanced relative to the non-perturbative
contributions improving convergence significantly and making pQCD evaluation
of exclusive form factors self-consistent toward lower values of $Q^2$, where
it may not be otherwise justified.
 
Such techniques of systematic organization of the potentially large
logarithmic contributions is a modification from the standard collinear 
factorization and is generally termed as the "{\it $k_T$-factorization}", 
that has been widely applied to inclusive as well as exclusive processes 
\cite{kT}.  However, unlike the familiar collinear factorization theorem, 
the  $k_T$-factorization is currently considered only at the level of a 
conjecture which is yet to be proven to all orders in perturbation theory 
(this is a highly debatable issue and, in fact, not yet fully recognized, 
e.g., see \cite{JPMa} for a different viewpoint.) To demonstrate that 
$k_T$-factorization is indeed a systematic tool, demands higher order 
calculations which may be very challenging. However, in 
this paper, we shall implicitly assume the validity of such a modified 
factorization without proving it and restrict ourselves at the tree level 
analysis of the $k_T$-dependent hard kernel. Moreover, in \cite{LiNagashima}, 
the $k_T$-factorization was proven at the level of twist-2 accuracy, while 
the collinear factorization was explicitely shown to be valid at the twist-3 
accuracy in the case of the $\pi\gamma^*\rightarrow\gamma$  transition form 
factor. Our analysis is, therefore, based on the key assumption that the same formalism could be straightforwardly 
extended to the elastic kaon form factors.

At the leading order $\sim1/Q^2$, the twist-2 and the two-particle 
twist-4 terms contribute to the hard kernels which have exactly the same 
expression given by
\begin{eqnarray}
\label{eq:amp2}
T^{\rm (t=2,4;LO)}_{\rm hard}(x,y,Q^2,{\mathbf k}_{T},{\mathbf l}_{T},\mu^2)=\frac{\pm16\pi\,
{\mathcal C}_F\,\alpha_s(\mu^2)\,x Q^2}{(xQ^2\pm{\mathbf k}_{T}^2)(xy\,Q^2\pm({\mathbf k}_{T}-{\mathbf l}_{T})^2)}\,,
\end{eqnarray}
while, the ${\mathcal{O}}(1/Q^4)$ power suppressed 2-particle twist-3 and  twist-4 hard 
kernels are, respectively, given by
\begin{eqnarray}
\label{eq:amp34}
T^{\rm(t=3)}_{\rm hard}(x,y,Q^2,{\mathbf k}_{T},{\mathbf l}_{T},\mu^2)
\!\!&=&\!\!\frac{32\pi\,
{\mathcal C}_F\,\alpha_s(\mu^2)\,x}{(xQ^2\pm{\mathbf
  k}_{T}^2)(xy\,Q^2\pm({\mathbf k}_{T}-{\mathbf l}_{T})^2)}\,;\nonumber\\
T^{\rm(t=4;NLO)}_{\rm hard}(x,y,Q^2,{\mathbf k}_{T},{\mathbf l}_{T},\mu^2)
\!\!&=&\!\!\frac{48\pi\,
{\mathcal C}_F\,\alpha_s(\mu^2)}{(xQ^2\pm{\mathbf
  k}_{T}^2)(xy\,Q^2\pm({\mathbf k}_{T}-{\mathbf l}_{T})^2)}\,.
\end{eqnarray}
In the above expressions, ${\mathcal C}_F=4/3$, the "$+$" signs correspond to the 
space-like case and the "$-$" signs correspond to the time-like case;
${\mathbf k}_T$ and ${\mathbf l}_{T}$ are, respectively, the initial and final 
relative transverse momenta of the valence quarks, and $x$ and $y$ are the 
corresponding light-cone momentum fractions. Note that the factors in the 
denominators that arise from the parton propagators develop poles in the 
time-like region.

To obtain the hard form factors, we use the following momentum 
space projection operator for the DAs with the different twist structures:
\begin{eqnarray}
{\mathcal M}^{K}_{\alpha\beta}\!&=&\!\frac{i}{4}\left\{P\!\!\!\!/\gamma_5\left(\Psi_{2;K}-\frac{1}{4}m^2_K\Psi^{\mathbb{A}}_{4;K}\partial^2 _{k_T}\right)+m^2_K\gamma_5\left(\frac{\bar{P}\!\!\!\!/}{\bar{P}\cdot P}\frac{\partial}{\partial x}\left(\int^x_0\Psi^{\mathbb{B}}_{4;K}\right)-\Psi^{\mathbb{A}}_{4;K}\partial_{k_T}\!\!\!\!\!\!\!\!/\,\,\,\,\right)\right.\nonumber\\
&&\!\!-\left.\mu_K\gamma_5\left(\Psi^p_{3;K}-\frac{i}{6}\sigma_{\mu\nu}n^\mu\bar{n}^\nu\frac{\partial}{\partial x}\Psi^{\sigma}_{3;K}+\frac{i}{6}\sigma_{\mu\nu}P^\mu\,\Psi^{\sigma}_{3;K}\,\partial^{\,\nu}_{k_T}\right)\right\}_{\alpha\beta}\,;\,\bar{P}=|P|\bar{n}\,,
\end{eqnarray} 
where $\mu_K=\frac{m^2_K}{m_u+m_s}$ GeV is the
"chiral-enhancement" parameter arising in the standard definition of the
2-particle twist-3 DAs (see, appendix), $\Psi^{\mathbb{B}}_{4;K}=\Psi^{g}_{4;K}-\Psi_{2;K}$,
$\partial_{k_T} \!\!\!\!\!\!\!\!/\,\,\,\,\equiv \gamma^{\mu}\partial/\partial
{k^{\mu}_T}$, $n=(1,0,{\mathbf 0}_T)$  the unit vector in the ``$+$'' direction, 
and $\bar{n}=(0,1,{\mathbf 0}_T)$ the unit vector in the ``$-$'' direction. 
Setting the renormalization/factorization scale to the magnitude of the 
incoming or outgoing kaon momentum i.e., $\mu=|P|=|P'|=Q/\sqrt{2}$ and 
convolving the projection operators for the kaon DAs with the hard kernels 
using factorization formula 
(symbolically, ${\mathcal M}^{\dagger}_{\rm K;out}\otimes T^{\rm LO}_{\rm
  hard} \otimes{\mathcal M}_{\rm K;in}$), we have
\begin{eqnarray}
\label{eq:factorization}
\left(P'\pm P\right)_{\mu}\!\!\!&&\!\!\!\!\!\!\{F,G\}^{\rm hard}_K(Q^2)\nonumber \\
&=&\!\!\int^1_0 dxdy\int\frac{d^2{\mathbf k}_{T}}{16\pi^3}
\,\frac{d^2{\mathbf l}_{T}}{16\pi^3} \,\left(\frac{4\pi\alpha_s(t)\,{\mathcal
    C}_F}{3}\right) {\rm exp}\left[-i{\mathbf k}_{T}\cdot{\mathbf b}_{1}-i{\mathbf l}_{T}\cdot{\mathbf b}_{2}\right]\nonumber\\
&&{\mathbf Tr}\!\left[\frac{\gamma^{\nu}{\mathcal
      M}_{K;out}^{\dagger}\,\gamma_{\nu}\,/\!\!\!k_{internal}\,\gamma_{\mu}\,{\mathcal
      M}_{K;in}}{(k^2_{internal}+i\epsilon)\,(k^2_g+i\epsilon)}+{\rm 3\,
    diagrams}\right]\,{\mathcal U}_{\rm RGE}(t,\mu)\nonumber \\
&&\times\,S_{t}(x)\,{\rm exp}\left[-S(x,y,|{\mathbf k}_{T}|\sim 1/b_1,|{\mathbf l}_{T}|\sim 1/b_2,\mu)\right]\,,
\end{eqnarray}
where $k_{internal}$ and $k_g$ are the internal quark and gluon momenta, 
respectively, as shown in Fig.1. Also, in the above equation
\begin{eqnarray}
{\mathcal U}_{\rm RGE}(t,\mu)={\rm exp}\left[4\int^{\mu}_{t}\frac{d\bar{\mu}}{\bar{\mu}}\gamma_q(\alpha_s({\bar{\mu}}^2))\right]\,;\,\gamma_q(\alpha_s({\bar{\mu}}^2))= -\frac{\alpha_s({\bar{\mu}}^2)}{\pi}\,,
\end{eqnarray}
represents the RG evolution factor for the scattering kernel from the 
``upper-factorization'' scale $t={\rm max}(\sqrt{x}\,Q,1/b_1,1/b_2)$ to 
the renormalization scale $\mu=Q/\sqrt{2}$, and $\gamma_q$ is the quark anomalous
dimension. The expression for the Sudakov exponent $S(x,y,b_1,b_2,Q)$ 
(after absorbing the RG factor from the kernel) is given by \cite{LiSterman}
\begin{eqnarray}
\label{eq:Sudakov_S}
S(x,y,b_1,b_2,Q)\!&=&\!s(xQ,b_1)+s((1-x)Q,b_1)+s(yQ,b_2)+s((1-y)Q,b_2)\nonumber\\
&&+2\int^{t}_{1/b_1}\frac{d\bar{\mu}}{\bar{\mu}}\gamma_q(\alpha_s({\bar{\mu}}^2))+2\int^{t}_{1/b_2}\frac{d\bar{\mu}}{\bar{\mu}}\gamma_q(\alpha_s({\bar{\mu}}^2))\,,
\end{eqnarray}
where
\begin{eqnarray}
\label{eq:Sudakov_s}
s(xQ,1/b)\equiv s(x\mu,1/b)=\int^{x\mu}_{1/b}\frac{d\bar{\mu}}{\bar{\mu}}\left[\ln\left(\frac{\mu}{\bar{\mu}}\right){\mathcal A}
(\alpha_s(\bar{\mu}^2))
  + {\mathcal B}(\alpha_s(\bar{\mu}^2))\right]\,, 
\end{eqnarray}
where the ``lower-factorization'' scales $1/b_1$, $1/b_2>\Lambda_{\rm QCD}$ 
serve to separate the perturbative from the non-perturbative transverse 
distances which are also typically the scales that provide a natural
starting point of the evolution of the kaon wavefunctions. In the above 
equations, the so-called ``cusp'' anomalous dimensions 
${\mathcal A}$ and ${\mathcal B}$, to one-loop accuracy are given by 
\begin{eqnarray}
{\mathcal A}(\alpha_s(\mu^2))\!&=&\!{\mathcal C}_F\frac{\alpha_s(\mu^2)}{
\pi}+\left[\left(\frac{67}{27}-\frac{\pi^2}{9}\right)N_c-\frac{10}{27}N_f+\frac{8}{3}\beta_0\ln
\left(\frac{e^{\gamma_E}}{2}\right)\right]\left(\frac{\alpha_s(\mu^2)}{\pi}\right)^2\,,\nonumber\\
{\mathcal B}(\alpha_s(\mu^2))\!&=&\!\frac{2}{3}\frac{\alpha_s(\mu^2)}{\pi}
\ln\left(\frac{e^{2\gamma_E-1}}{2}\right)\,.
\end{eqnarray}
The  exact form of the threshold resummation ``jet'' function $S_t(x)$ in 
Eq.~\ref{eq:factorization} involves a one parameter integration, but in 
practice it is more convenient to take the simple parameterization 
proposed in \cite{Sanda,Li}:
\begin{equation}
 S_t(x)=\frac{2^{1+2c}\Gamma(3/2+c)}{\sqrt{\pi}\Gamma(1+c)}\left[x(1-x)\right]^c\,,
\end{equation}
where the parameter $c\approx 0.3$ for light pseudo-scalar mesons like the pion
and the kaon.

Now we present the factorized result for the hard kaon form factors up to 
twist-4 corrections as follows: 
\begin{eqnarray}
\{F,G\}^{\rm hard}_K(Q^2)\!\!&=&\!\!\delta\{F,G\}^{\rm twist2}_K(Q^2)+\delta\{F,G\}^{\rm twist3}_K(Q^2)+\delta\{F,G\}^{\rm twist4}_K(Q^2)\,;\nonumber\\
\delta\{F,G\}^{\rm twist4}_K(Q^2)\!\!&=&\!\!\delta\{F,G\}^{\rm twist4;LO}_K(Q^2)+\delta\{F,G\}^{\rm twist4;NLO}_K(Q^2)\,,
\end{eqnarray}
where the leading twist-2 and twist-4 corrections are expressed by the 
following integral representations in the impact parameter space:
\begin{eqnarray}
\label{eq:twist24FF}
\delta\{F,G\}^{\rm twist2}_K\!\!&&\!\!\!(Q^2)+\delta\{F,G\}^{\rm twist4;LO}_K(Q^2)=32\pi Q^2{\mathcal C}_F\int^1_0 x\,dxdy\int^{\infty}_{0} b_1db_1b_2 db_2\,\alpha_s(t)\nonumber\\
&&\hspace{-0.0cm}\times\,\left[\pm\frac{1}{2}\,{\mathcal P}_{2;K}(x,b_1)\,{\mathcal P}_{2;K}(y,b_2)\mp\, m^2_K\,\frac{b^2_2}{8}\,{\mathcal P}_{2;K}(x,b_1)\,{\mathcal P}^{\mathbb{A}}_{4;K}(y,b_2)\right.\nonumber\\
&&\hspace{-0.0cm}\left.\mp \,m^2_K\,\frac{b^2_1}{8}\,{\mathcal P}^{\mathbb{A}}_{4;K}(x,b_1)\,{\mathcal P}_{2;K}(y,b_2)+{\mathcal{O}}\left(m^4_K b^4_1,m^4_K b^4_2\right)\right]\nonumber\\
&&\hspace{-0.0cm}\times\,H_{\pm}(x,y,Q,b_1,b_2)\,{\rm exp}\left[-S(x,y,b_1,b_2,Q)\right]\,S_{t}(x)\,,
\end{eqnarray}
while, the power suppressed twist-3 and twist-4 corrections are given by
\begin{eqnarray}
\label{eq:twist34FF}
\delta\{F,G\}^{\rm twist3}_K\!\!\!\!&&\!\!(Q^2)+\delta\{F,G\}^{\rm twist4;NLO}_K(Q^2)=32\pi Q^2{\mathcal C}_F\int^1_0 dxdy\int^{\infty}_{0}b_1db_1b_2db_2\,\alpha_s(t)\nonumber\\
&&\hspace{-0cm}\times\,\left[\frac{\mu^2_K}{Q^2}\left({\bar x}\,{\mathcal P}^{\,p}_{3;K}(x,b_1)\,{\mathcal P}^{\,p}_{3;K}(y,b_2)+\frac{(1+x)}{6}\,\frac{\partial}{\partial x}{\mathcal P}^{\,\sigma}_{3;K}(x,b_1)\,{\mathcal P}^{\,p}_{3;K}(y,b_2)\right.\right.\nonumber\\
&&\hspace{-0cm}\left.\left.+\,\frac{1}{2}{\mathcal P}^{\,\sigma}_{3;K}(x,b_1)\,{\mathcal P}^{\,p}_{3;K}(y,b_2)\right)+\frac{3m^2_K}{2Q^2}\left(\int^x_0d\zeta\,{\mathcal P}^{\mathbb{B}}_{4;K}(\zeta,b_1)\right)\right.\nonumber\\
&&\hspace{-0cm}\left.\times\,\left({\mathcal P}_{2;K}(y,b_2)-\,m^2_K\,\frac{b^2_2}{4}\,{\mathcal P}^{\mathbb{A}}_{4;K}(y,b_2)\right)+{\mathcal{O}}\left(m^4_K b^4_1,m^4_K b^4_2\right)\right]\nonumber\\
&&\hspace{-0cm}\times\,\,H_{\pm}(x,y,Q,b_1,b_2)\,{\rm exp}[-S(x,y,b_1,b_2,Q)]\,S_{t}(x)\,,
\end{eqnarray}
where ${\mathcal P}_{t;K}(x,b)\equiv\tilde{\mathcal P}_{t;K}(x,b,1/b^2,{\mathcal{M}}_{\{u,d,s\}})$, 
and ${\mathcal P}^{\mathbb{B}}_{4;K}\!=\!{\mathcal P}^{g}_{4;K}-{\mathcal P}_{2;K}$.
Note that the ``LO'' and ``NLO'' used in the above equation should not be
confused with the usual terminologies associated with perturbative expansions
in terms of $\alpha_s$ but rather in the sense of operator product expansion 
(OPE) terms. In the impact representation, the space- and time-like hard
kernels (the part of the scattering kernel that is common to all the twists) 
could be expressed in terms of standard Bessel functions $K_0, \,I_0,\, J_0$ 
and $H^{(1)}_0$ and are given by 
\begin{eqnarray}
H_{+}(x,y,Q,b_1,b_2)\!&=&\!K_0(\sqrt{xy}\,Qb_2)\nonumber\\
&&\times\left[\theta(b_1-b_2)K_0(\sqrt{x}\,Qb_1)I_0(\sqrt{x}\,Qb_2)\right.\nonumber\\
&&+\left. \theta(b_2-b_1)K_0(\sqrt{x}\,Qb_2)I_0(\sqrt{x}\,Qb_1)\right]\,;
\end{eqnarray}
\begin{eqnarray}
H_{-}(x,y,Q,b_1,b_2)\!\!&=&\!\!\left(\frac{i\pi}{2}\right)^2H^{(1)}_0(\sqrt{xy}\,Qb_2)\nonumber\\
&&\times\left[\theta(b_1-b_2)H^{(1)}_0(\sqrt{x}\,Qb_1)J_0(\sqrt{x}\,Qb_2)\right.\nonumber\\
&&+\left. \theta(b_2-b_1)H^{(1)}_0(\sqrt{x}\,Qb_2)J_0(\sqrt{x}\,Qb_1)\right]\,,
\end{eqnarray}
where $H_{+}$ is a real-valued function and $H_{-}$ is a complex-valued
function of real arguments. 

{\it Apropos} of our derived formulas Eqs.~\ref{eq:twist24FF} and 
\ref{eq:twist34FF}, it is noteworthy to mention that in \cite{Pire} 
it was suggested that the Sudakov factors must be analytically continued 
from the space-like to the time-like case. This may not be generally true. 
The Sudakov factors in \cite{Magnea} (see, section 3.1 of this reference) 
arise directly from ``form factor-type'' kernels, which are not universal 
quantities and may vary with processes. There the analytic continuation is 
perfectly justified. However, for an approach based on the factorization 
theorem, one uses ``universal'' Sudakov factors $S(Q)$ arising from the 
overlap of the soft and collinear processes below the factorization scale, 
as in the present context. As explained in \cite{Coriano}, these Sudakov 
factors are to be considered as an integral part of the DAs and, thus,
they are universal quantities as well, depending only on the magnitude of 
the energy scale $Q\geq 0$. Note that the $Q$ dependence of the Sudakov 
factor in Eqs.~\ref{eq:Sudakov_S} and \ref{eq:Sudakov_s}, stems from the 
dependence on the collinear components of the external pion 4-momenta which 
are given by $P^+=P'^-=Q/\sqrt{2}$, in the Breit-frame. As such, it is 
important that one does not analytically continue but rather use the 
same Sudakov factor in both the space and time-like cases.

\subsection{Non-factorizable Soft QCD}
In \cite{Nest} it was shown that the space-like low-energy pion data below 
$Q^2\sim10$ GeV$^2$ is dominated by the soft pion form factor which 
accounts for more that 70\% of the data. Such soft QCD contributions are 
non-factorizable and are beyond the realm of ordinary perturbation theory. 
Since, no systematic method is currently available to calculate these 
non-perturbative effects, one is compelled to use some model ansatz
to obtain a rough estimate of their contributions, viz, in \cite{Nest}
the soft pion form factor in the space-like region was calculated using the 
{\it Local Duality} (LD) model in QCDSR. In our present work, we extend the
same result to the space-like kaon form factor which is then given by
\begin{equation}
\label{eq:soft_s}
\left.F^{\rm soft}_K(Q^2)\right|_{\rm LD}=1-\frac{1+6s_0/Q^2}
{(1+4s_0/Q^2)^{3/2}}\approx \frac{6s_0}{Q^4}+{\mathcal
  O}(\frac{1}{Q^6})\,;\,\,s_0\approx 4\pi^2 f^2_K\,.
\end{equation} 
Now, on one hand, an {\it ab initio} derivation of the corresponding time-like
soft form factor seems {\it a priori} unfeasible using  QCDSR, since the LD 
principle is strictly applicable for the space-like region only.  
On the other hand, a naive analytic continuation of the space-like formula, 
i.e., by a replacement of $Q^2\rightarrow -Q^2$, leads to an undesirable
pole in the denominator of the soft form factor:
\begin{eqnarray}
\label{eq:soft_an}
G^{\rm soft}_{K;{\rm analytic}}(Q^2)\Rightarrow 1-\frac{1-6s_0/Q^2}{(1-4s_0/Q^2)^{3/2}}\approx
\frac{6s_0}{Q^4}+{\mathcal O}(\frac{1}{Q^6})\,.
\end{eqnarray}  
Since, here our primary goal is to
obtain an estimate for the smooth continuum part of the kaon spectra for 
intermediate energies which in reality is, however, dominated by low-energy 
time-like resonances that obscure the smooth continuum. With the 
"over-simplified" assumption that these resonance peaks behave as background 
"noise", superimposed on a smooth continuum spectrum, we choose the functional 
form of the time-like soft form factor to be the same as that of the
space-like expression, which is a smooth function for the entire range of 
$Q^2$ we consider, i.e.,
\begin{equation}
\label{eq:soft_t}
G^{\rm soft}_K(Q^2)=1-\frac{1+6s_0/Q^2}
{(1+4s_0/Q^2)^{3/2}}+{\mathcal
  O}(\frac{1}{Q^6})\,.
\end{equation}
Moreover, for large enough $Q^2\sim$ above 5 GeV$^2$, both expressions 
Eqs.~\ref{eq:soft_s} and \ref{eq:soft_an} when expanded in inverse powers 
of $Q^2$ yield the same leading term of ${\mathcal O}(\frac{1}{Q^4})$. 
Hence, the particular choice of the soft form factors should not matter
significantly at large-$Q^2$ values where the perturbative predictions 
become more reliable and dominant.

In the present context, a vital aspect deserves some consideration. Since, the 
inclusion of the soft form factors has been somewhat {\it ad hoc}, without any 
correspondence among the hard and the soft contributions, there could be
chances of possible double-counting of contributions especially at low
energies. Thus, it becomes clear that we must correct the hard factorized 
results in the low-$Q^2$ region to ensure that the respective contributions 
lie within their domains of validity. This is achieved by enforcing the gauge 
invariance condition through the {\it vector Ward-identity} $\{F,\,G\}_K(Q^2=0)=1$, 
which is {\it a priori} not ensured in perturbative calculations.
Since the soft form factors satisfies $\{F,G\}^{\rm soft}_K(Q^2=0)=1$, we must have 
$\{F,G\}^{\rm hard}_K(Q^2=0)=0$. But this is unfortunately not satisfied by
Eqs.~\ref{eq:twist24FF} and \ref{eq:twist34FF} where the contributions tend to
diverge rapidly in the vicinity of $Q^2=0$. Therefore, the essential task is to match
the large-$Q^2$ results of $\{F,G\}^{\rm hard}_K(Q^2)$ with the low-$Q^2$ results 
of $\{F,G\}^{\rm soft}_K(Q^2)$. Here we shall modify the argument given in \cite{Bakulev} 
for the twist-2 case to be applicable for the twist-3 and twist-4 power corrections. 
The simplest way is to ``power-correct'' for the singular $\sim 1/Q^2$ (leading 
twist-2 and twist-4) and $\sim 1/Q^4$ (sub-leading twist-3 and twist-4) behaviors, 
respectively, at small $Q$, by introducing some characteristic low-energy mass scale 
$M_0$ that may lead to the onset of the genuine non-perturbative soft dynamics. 
For the soft form factors modeled via LD principle, the scale $M^2_0=2s_0$ is a natural
choice \cite{BakRad}. It can then be shown that for the leading twist-2 and
twist-4 hard corrections, it is sufficient to make the modification \cite{Bakulev}
\begin{eqnarray}
\delta\{F,G\}^{\rm twist2}_K(Q^2)\!\!&+&\!\!\delta\{F,G\}^{\rm
  twist4;LO}_K(Q^2)\nonumber\\
&&\hspace{-2.5cm}\rightarrow\Delta\{F,G\}^{\rm twist2}_K(Q^2)+\Delta\{F,G\}^{\rm
  twist4;LO}_K(Q^2)\nonumber\\
&&\hspace{-2.5cm}=\left(\frac{Q^2}{2s_0+Q^2}\right)^2
\left(\delta\{F,G\}^{\rm twist2}_K(Q^2)+\delta\{F,G\}^{\rm
  twist4;LO}_K(Q^2)\right)\,.
\end{eqnarray}
For the case of the sub-leading twist-3 and twist-4 hard corrections, 
we perform the following replacement:
\begin{eqnarray}
\delta\{F,G\}^{(t=3,4)}_K(Q^2)=\widetilde{\delta\{F,G\}}^{(t=3,4)}_K\hspace{-0.4cm}(Q^2)
\frac{M^4_0}{Q^4}\rightarrow\widetilde{\delta\{F,G\}}^{(t=3,4)}_K\hspace{-0.4cm}(Q^2)\frac{M^4_0}{M^4_0+Q^4}\,,
\end{eqnarray}
where we write $\delta\{F,G\}^{(t=3,4)}_K\equiv\delta\{F,G\}^{\rm
  twist3}_K+\delta\{F,G\}^{\rm twist4;NLO}_K$ for brevity.
Now, to maintain the Ward-identity, we correct for the wrong $Q^2=0$ 
limit of the above expression
\begin{eqnarray}
\Delta\{F,G\}^{(t=3,4)}_K(Q^2)=-\widetilde{\delta\{F,G\}}^{(t=3,4)}_K\hspace{-0.4cm}(Q^2)\,
\Phi_n(Q^2/M^2_0)+\widetilde{\delta\{F,G\}}^{(t=3,4)}_K\hspace{-0.4cm}(Q^2)\frac{M^4_0}{M^4_0+Q^4}\,,
\end{eqnarray}
where we introduce the smooth function $\Phi_n(z)$ (with $z=Q^2/M^2_0$) with the 
essential property that $\Phi_n(0)=1$  and $z\,\Phi_n(z)\rightarrow0$ as 
$z\rightarrow\infty$, for a suitable choice of the positive integer $n$, 
to preserve the asymptotics of $\delta\{F,G\}^{(t=3,4)}_K(Q^2)$. A natural 
choice for $\Phi_n(z)$ could be $\Phi_n(z)=1/(1+z^n)^2$, concurrent with
the $\sim 1/Q^{2n}$ scaling behavior of the respective power suppressed
terms. For the present purpose, it is sufficient to take $n=2$, yielding
\begin{eqnarray}
\Delta\{F,G\}^{(t=3,4)}_K(Q^2)\!&=&\!\widetilde{\delta\{F,G\}}^{(t=3,4)}_K\hspace{-0.4cm}(Q^2)\,
\frac{M^4_0}{M^4_0+Q^4}\left(1-\frac{M^4_0}{M^4_0+Q^4}\right)\nonumber\\
&=&\!\delta\{F,G\}^{(t=3,4)}_K(Q^2)\,\left(\frac{Q^4}{M^4_0+Q^4}\right)^2\,.
\end{eqnarray}
In principle, this can also be achieved with larger integer values of $n$ that would lead
to $\left(Q^{2n}/(M^{2n}_0+Q^{2n})\right)^2$ in front of the hard
parts. However, as $n\rightarrow\infty$, this factor becomes a step function,
which is no longer smooth. Hence, a minimal value of $n$ is preferable
and we arrive at the Ward-identity modified result:
\begin{eqnarray}
\delta\{F,G\}^{\rm twist3}_K(Q^2)\!\!&+&\!\!\delta\{F,G\}^{\rm
  twist4;NLO}_K(Q^2) \nonumber\\
&&\hspace{-2.5cm}\rightarrow\Delta\{F,G\}^{\rm twist3}_K(Q^2)+\Delta\{F,G\}^{\rm
  twist4;NLO}_K(Q^2) \nonumber\\
&&\hspace{-2.5cm}=\left(\frac{Q^4}{4s^2_0+Q^4}\right)^2\left(\delta\{F,G\}^{\rm
  twist3}_K(Q^2)
+\delta\{F,G\}^{\rm twist4;NLO}_K(Q^2)\right)\,.
\end{eqnarray} 
The pre-factors only alter the low-energy behavior of 
the hard contributions and ensure the correct power-law in maintaining 
a smooth matching between the large-$Q^2$ behavior of 
$\{F,G\}^{\rm hard}_K(Q^2)$ to the low-$Q^2$ behavior of 
$\{F,G\}^{\rm soft}_K(Q^2)$ (see, Fig.~\ref{fig:relFF} ). This leads to 
our final expression for the total electromagnetic kaon form factors, 
correct up to ${\mathcal O}(\frac{1}{Q^4})$ accuracy, given by
\begin{equation}
\{F,G\}_K(Q^2)\!=\!\{F,G\}^{\rm soft}_K(Q^2)+\Delta\{F,G\}^{\rm twist2}_K(Q^2)+\Delta\{F,G\}^{\rm
  twist3}_K(Q^2)+\Delta\{F,G\}^{\rm twist4}_K(Q^2)\nonumber\,,
\end{equation}
where
\begin{eqnarray}
\label{eq:final_FF}
\{F,G\}^{\rm soft}_K(Q^2)\!\!&=&\!\!1-\frac{1+6s_0/Q^2}{(1+4s_0/Q^2)^{3/2}}\,,\nonumber\\
\Delta\{F,G\}^{\rm twist2}_K(Q^2)\!\!&=&\!\!\left(\frac{Q^2}{2s_0+Q^2}\right)^2\,\delta\{F,G\}^{\rm twist2}_K(Q^2)\,,\nonumber\\
\Delta\{F,G\}^{\rm
  twist3}_K(Q^2)\!\!&=&\!\!\left(\frac{Q^4}{4s^2_0+Q^4}\right)^2\,\delta\{F,G\}^{\rm
  twist3}_K(Q^2)\,,\nonumber\\
\Delta\{F,G\}^{\rm
  twist4}_K(Q^2)\!\!&=&\!\!\left(\frac{Q^2}{2s_0+Q^2}\right)^2\,\delta\{F,G\}^{\rm
  twist4;LO}_K(Q^2)\nonumber\\
&+&\!\!\left(\frac{Q^4}{4s^2_0+Q^4}\right)^2\,\delta\{F,G\}^{\rm
  twist4;NLO}_K(Q^2)\,,
\end{eqnarray}
where the $\delta$'s are replaced by the $\Delta$'s to include the
respective pre-factors.
\section{Results and Discussion}
To obtain the BHL Gaussian parameters of the kaon wavefunctions, we use 
the following two sets of constraints valid for the individual twists 
($t=2,3,4$): the first set of constraints is obtained from the leptonic 
decay $K\rightarrow\mu+\nu_{\mu}$, and given by
\begin{equation}
\int^1_0 dx\,\int \frac{d^2{\mathbf k}_{T}}{16\pi^3}\,\Psi_{t;K}(x,{\mathbf k}_T,\mu^2_0,{\mathcal M}_{\{u,d,s\}})=N_{t;K}\,,
\end{equation} 
with $N_{t;K}$ being the normalization constant for the collinear DAs; 
and the second follows from the phenomenological fact that the average 
transverse momentum of the valence partons in light mesons is about 
$\left<{\mathbf k}^2_T\right>^{1/2}_{\pi,K,\eta\cdots}\approx0.35$ GeV, i.e.,
\begin{equation}
\left<{\mathbf k}^2_T\right>_K=\frac{\int dx \int d^2{\mathbf k}_{T}
  \,\left|{\mathbf k}^2_{T}\right| \left|\Psi_{t;K}(x,{\mathbf k}_{T},\mu^2_0,{\mathcal
    M}_{\{u,d,s\}})\right|^2}{\int dx \int d^2{\mathbf k}_{T} \,\left|\Psi_{t;K}(x,{\mathbf k}_{T},\mu^2_0,{\mathcal M}_{\{u,d,s\}})\right|^2}\,.
\end{equation}
The Gaussian parameters determined in this way for $\mu_0\approx1$ GeV 
are collected in Table.~\ref{table:para-det}. Note that due to the rather 
mild scale dependences of these parameters, which practically remain constant 
for the entire range of intermediate energies that is considered in this work,
their scale variations have been kept fixed to reduce the numerical
complexity. However, we do consider their variation with the changes in the
collinear DA parameters, summarized in Table.~\ref{table:para}, that is required for
our estimation of the theoretical error. Once all the phenomenological 
parameters are determined, we proceed to calculate the hard contributions. 
For calculations, we use the full non-asymptotic collinear DAs 
derived from LCSR \cite{Filyanov,Ball,Lenz}.   

\begin{table}[t]
\begin{center} 
\begin{tabular}{|c|c|c||c|c|c|}
\hline
$K^\pm $ parameters & At $\mu_0=1$ GeV & units & $K^\pm $ parameters & At
$\mu_0=1$ GeV & units \\
\hline\hline
$m_{u,d}$ & $5.6\pm 1.6$\cite{Lenz} & MeV & $a^K_2$ & $0.25\pm 0.15$\cite{Lenz} & - \\
$m_s$ & $137\pm 27$\cite{Lenz} & MeV & $f_{K}$ & $1.22f_\pi$, $f_\pi=131$\cite{CZ} & MeV\\
${\mathcal M}_{u,d}$ & $0.33$ & GeV & $f_{3K}$ & $0.0045\pm 0.0015$\cite{Lenz} & GeV$^2$ \\ 
${\mathcal M}_{s}$ & $0.45$ & GeV  & $\omega_{3K}$ & $-1.2\pm 0.7$\cite{Lenz} & -\\ 
$m_K$         & $493$ & MeV & $\delta^2_{K}$& $0.20\pm 0.06$\cite{Lenz} &  GeV$^2$ \\ 
$a^K_1$       & $0.06\pm 0.03$\cite{Lenz} & - & $\omega_{4K}$& $0.2\pm 0.1$\cite{Lenz} & - \\ 
\hline
\end{tabular}
\end{center}
\caption{Various input parameters for twist-2, twist-3 and twist-4 
wavefunctions.}
\label{table:para}
\end{table}
\begin{table}[t]
\begin{center} 
\begin{tabular}{|c|c|c||c|c|c|}
\hline
$A_{t;K} (t=2,3,4) $ &  At $\mu_0=1$ GeV  & units & $(\beta_{t;K})^2 (t=2,3,4) $
&  At $\mu_0=1$ GeV &  units   \\
\hline\hline
$A_{2;K}$ & $2.06\,(2.07)_{\rm as}$ & - & $(\beta_{2;K})^2$  &
$0.78\,(0.89)_{\rm as}$ & GeV$^{-2}$   \\ 
$A^{p}_{3;K}$ & $2.23\,(2.28)_{\rm as}$ & - & $(\beta^{\,p}_{3;K})^2$ &
$0.71\,(0.79)_{\rm as}$ & GeV$^{-2}$ \\ 
$A^{\sigma}_{3;K}$ & $2.08\,(2.07)_{\rm as}$ & - &
$(\beta^{\,\sigma}_{3;K})^2$ & $0.88\,(0.89)_{\rm as}$ & GeV$^{-2}$ \\ 
$A^{\mathbb{A}}_{4;K}$ & $2.07\,(2.00)_{\rm as}$ & - &
$(\beta^{\,\mathbb{A}}_{4;K})^2$ & $0.91\,(0.93)_{\rm as}$ & GeV$^{-2}$ \\ 
$A^{g}_{4;K}$ & $5.22\,(2.28)_{\rm as}$ & - & $(\beta^{\,g}_{4;K})^2$ &
$0.57\,(0.79)_{\rm as}$ & GeV$^{-2}$ \\ 
\hline
\end{tabular}
\end{center}
\caption{The phenomenologically determined Gaussian parameters 
        for twist-2, twist-3 and twist-4 wavefunctions. 
        The numbers in the parentheses $(\cdots)_{\rm as}$ 
        correspond to parameters for the asymptotic wavefunctions.}
\label{table:para-det} 
\end{table}
\begin{figure}
    \hspace{-0.6cm}\begin{tabular}{cc}
      \resizebox{13cm}{!}{\includegraphics{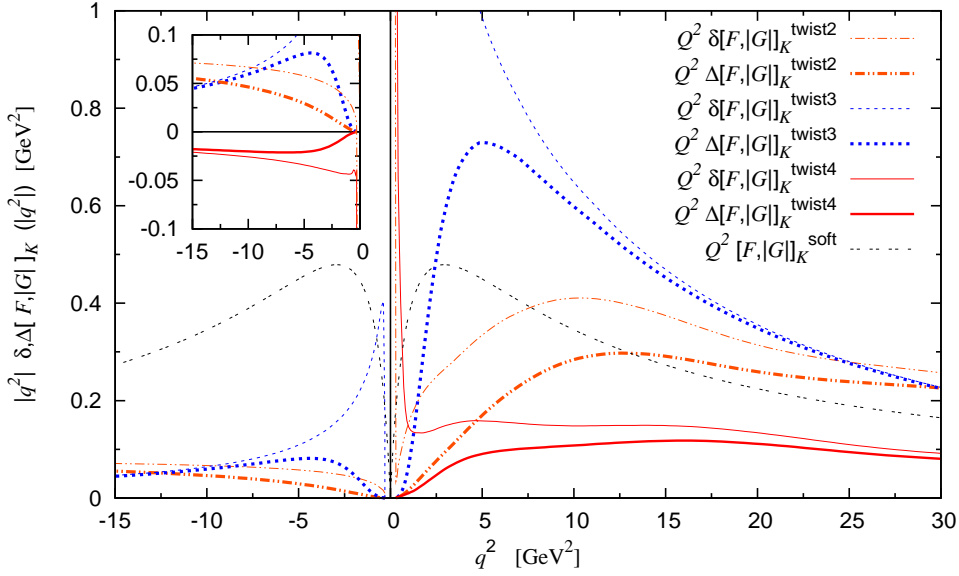}} 
    \end{tabular}
    \caption{Relative contributions of the soft $\{F,G\}^{\rm soft}_K$
      (double-dot black lines), twist-2 $\Delta\{F,G\}^{\rm twist2}_K$ (thick double-dot
      dashed orange lines), twist-3 $\Delta\{F,G\}^{\rm twist3}_K$ (thick
      dotted blue lines), and twist-4 $\Delta\{F,G\}^{\rm twist4}_K$ (thick
      solid red lines) terms in Eq.~\ref{eq:final_FF}. The same terms without 
      the pre-factor modifications are also displayed.}
    \label{fig:relFF}
\end{figure}
 
\begin{figure}
    \hspace{-0.5cm}\begin{tabular}{cc}
      \resizebox{8.2cm}{!}{\includegraphics{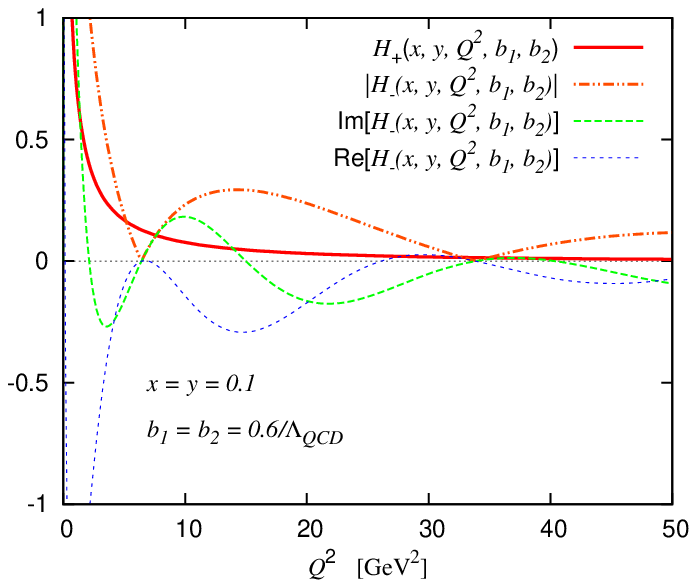}} &
      \resizebox{8.2cm}{!}{\includegraphics{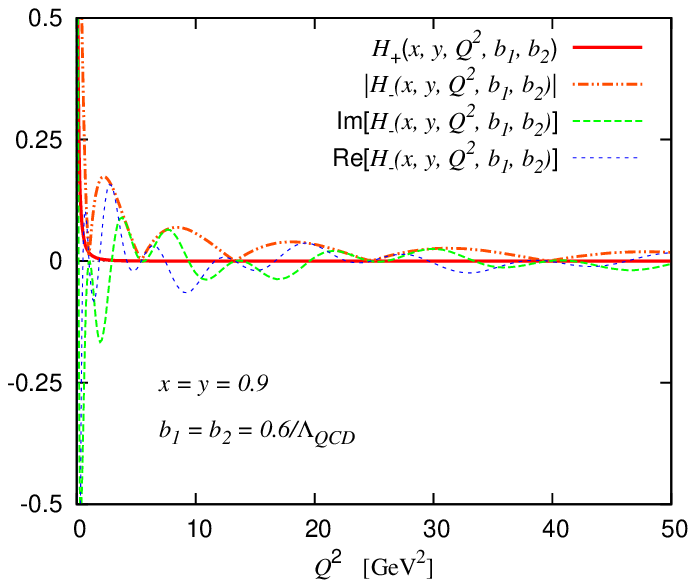}} \\
    \end{tabular}
    \caption{The space- and time-like hard kernels $H_\pm(x,y,Q,b_1,b_2)$ in
      the impact space representation for two sets of choices for the collinear
      momentum fractions with arbitrary fixed $b_1,b_2$: $x=y=0.1$ (left plot) and
      $x=y=0.9$ (right plot).}
    \label{fig:kernel}
\end{figure}

In the Fig.~\ref{fig:relFF}, we plot the individual terms of Eq.~\ref{eq:final_FF}, 
i.e., $\{F,G\}^{\rm soft}_K$, $\Delta\{F,G\}^{\rm twist2}_K$, 
$\Delta\{F,G\}^{\rm twist3}_K$ and $\Delta\{F,G\}^{\rm twist4}_K$, which should
give an idea about the relative magnitude of each contribution for intermediate 
values of $Q^2$ up to 30 GeV$^2$. For comparison, we also display the results 
obtained without including the pre-factor modifications, which do not show any
appreciable difference for $Q^2$ values beyond $\sim 5-10$ GeV$^2$. As expected, 
the standard twist-2 contributions are much smaller compared to the soft QCD and
the twist-3 power corrections at moderate-energies. However, the twist-4 
contributions are seen to be indeed small (about 1/3 of the magnitude of the
twist-2), which are, in fact, negative in the space-like region. In the
time-like region, since all the hard contributions are complex, it only 
makes sense to plot the modulus of the individual twist corrections. It is 
notable that the general enhancement of all the time-like hard contributions 
relative to the space-like ones can be attributed to the time-like parton propagators 
developing poles that are absent in the space-like region. To illustrate this 
point, it is useful to plot the part of the hard kernel $H_\pm(x,y,Q,b_1,b_2)$ 
that is common to all the twist corrections to the hard form factors. Fig.~\ref{fig:kernel}
shows the variation of the space- and time-like kernels $H_\pm$ (in the impact 
parameter space) as a function of $Q^2$ for some arbitrary fixed values of the 
parameters $x, y, b_1$ and $b_2$. It immediately becomes clear that the 
real-valued space-like kernel $H_+$ has a rapidly decaying exponential behavior, 
whereas the complex-valued time-like kernel $H_-$ has rather large amplitude 
oscillatory real and imaginary components which decay very gradually with
increasing $Q^2$. In reference to Eqs.~\ref{eq:amp2} and \ref{eq:amp34}, we note that if 
$\{x,y\}\ll 1 \Rightarrow \{xQ^2,\,xyQ^2\}\sim \{{\mathbf k}^2_T,\,{\mathbf l}^2_T\}\ll
Q^2$, and if $\{x,y\}\sim 1 \Rightarrow \{xQ^2,\,xyQ^2\}\sim Q^2$, so that the
terms in the denominators tend to cancel each other in the time-like but
not in the space-like domain. This explains why the amplitude of the time-like 
oscillations in $H_-$ grow larger and larger near the endpoints $x,y\rightarrow 0$. 

The most striking feature of our results in Fig.~\ref{fig:relFF} is the
anomalously large twist-3 contribution in the time-like region, similar to what 
was seen for the pion \cite{Raha2}, dominating all the other corrections 
for the entire range of low- and moderate-energies. This huge asymmetry 
between the space- and time-like twist-3 contributions comes from the additional 
parametric enhancement of the twist-3 DAs due to the chiral parameter $\mu_K$ 
which makes them particularly sensitive to the chiral scale. It is the combination 
of this parametric enhancement along with the occurrence of the time-like poles in 
the hard kernel that leads to such a characteristic anomalous twist-3 behavior
which is completely missing in the twist-2 or even in the twist-4. At this 
point, one may also worry about possible large contributions from the
3-particle twist-3 sector (related to the 2-particle twist-3 sector through
QCD equations of motion) that was not considered in this work. Here, we note 
that such a possibility can safely be precluded since the 3-particle twist-3 
DA receives large parametric suppression from the non-perturbative parameter 
$f_{3K}\approx0.0045$ GeV$^2$, numerically very much smaller compared 
to the analogous 2-particle twist-3 parameter $\mu_K\approx 1.5$ GeV which 
greatly enhances the contribution from the 2-particle sector. 

Further, it is important to note that (a) the 
``active'' soft gluons that may also arise from the 3-particle twist-3 DA or 
higher twist DAs likewise, bring about additional power corrections and, 
therefore, can be safely neglected at large-$Q^2$ values, and (b) the
``long distance'' soft gluons that may be a possible source of the  breakdown 
of TMD factorization, can not probe the small ``color-dipole'' configurations 
of the $q\bar{q}$ hadronic bound state at high enough $Q^2$ (color-transparency). 
The remaining collinear gluons are assumed to be effectively tackled within the
present TMD factorization scheme, where the inclusion of the 2-particle twist-3 
corrections indeed turn out to be the most crucial aspect at the moderate-$Q^2$ 
regime. Note, however, that all such non-perturbative power corrections 
including the soft contributions rapidly fall-off with increasing $Q^2$, and beyond 
$\sim 50-100$ GeV$^2$ the standard twist-2 terms start dominating the 
asymptotic regime, yielding back numerically the {\it bona fide} asymptotic behavior
given by the {\it Farrar and Jackson} result \cite{FJ}. 
\begin{equation}
\label{eq:FJ}
\{F,G\}^{\rm asy}_K(Q^2)=\frac{8\pi\alpha_s(Q^2)f^2_K}{Q^2}\,.
\end{equation}
\begin{figure}
    \hspace{-0cm}\begin{tabular}{cc}
      \resizebox{14cm}{!}{\includegraphics{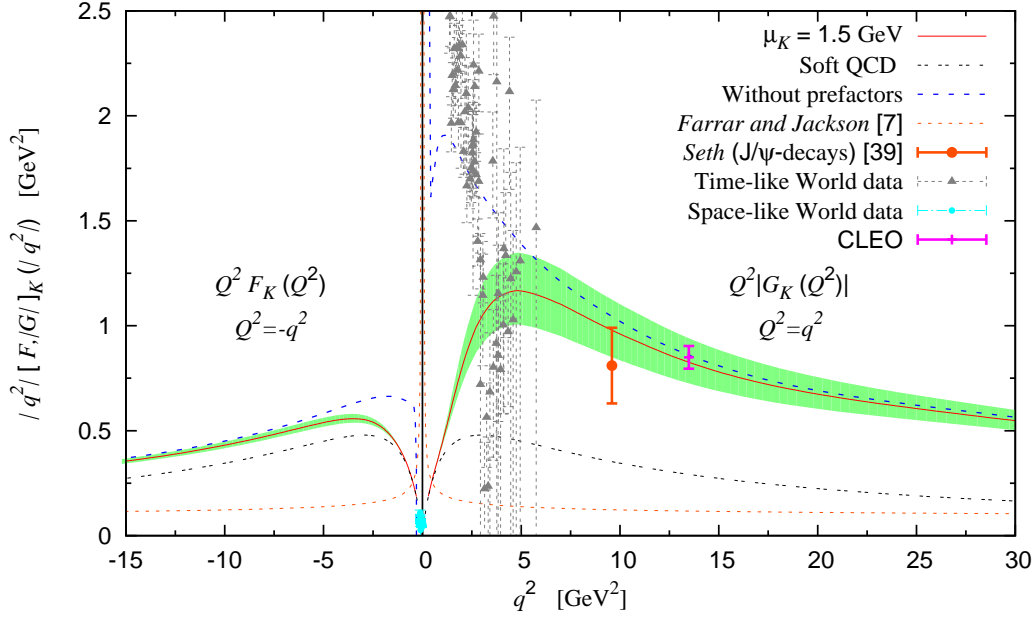}} 
    \end{tabular}
    \caption{The total scaled kaon form factors, denoted by the thick solid 
    (red) lines; the soft form factors $\{F,G\}^{\rm soft}_K$, denoted by the 
    double-dot (black) lines; and the asymptotic QCD result \cite{FJ}, 
    denoted by the short-dashed (orange) lines. The shaded area is our estimated 
    theoretical error. The experimental data taken from \cite{ee-hh,CLEO} 
    and the phenomenological result \cite{Seth} is shown for comparison.} 
    \label{fig:total_FF}
\end{figure}  
Our final prediction for the total scaled electromagnetic kaon form factors 
$\{F,G\}_K$  (from Eq.~\ref{eq:final_FF}) up to twist-4 accuracy in the range 
of intermediate energies/momentum transfers is presented in
Fig.~\ref{fig:total_FF}, along with the result for the soft form 
factors $\{F,G\}^{\rm soft}_K$ and the standard asymptotic QCD result of 
{\it Farrar and Jackson} \cite{FJ} for comparison. To estimate the theoretical
error we studied the variation of the wavefunction parameters provided in the 
Tables~\ref{table:para} and \ref{table:para-det}. In addition, we varied
the chiral parameter $\mu_K=m^2_K/(m_u+m_s)$ which is often taken to be
slightly lower $\sim 1.3-1.5$ GeV in the literature \cite{Raha1,Raha2,Keum,Sanda,Li,Lu,Ball,B-decay,CHPT} 
than its naive value about 1.7 GeV expressed in terms of the current quark
masses. In this analysis, we take $\mu_K=1.5\pm 0.2$ and include its variation 
in the error estimate. The shaded area, thus obtained, can be regarded as our
rough estimate for the theoretical error, where the solid (red) curve corresponds 
to the central values of the parameters. While our result is relatively insensitive 
to the choice of the parameters in the space-like region, the time-like 
result turns out to be very sensitive to the choice of $\mu_K$ whose variation
alone amounts for more than $90\%$ of the error-bar. The error due to
the rest of the model parameters is generously overestimated to include
possible uncertainties due to the soft parts which we do not {\it a priori}\, 
take into account. Thus, we should stress that our pQCD based error estimate 
in the low-$Q^2$ region (which apparently looks small) must be considered in 
a very conservative sense and can not be taken seriously below $\sim 5$
GeV$^2$. A more rigorous error analysis is impossible at the moment due to 
poor quality of the experimental data.
   
Several comments are now in order:   
\begin{itemize}
\item The width of our error-bar is large enough to completely subsume effects
  due to further inclusion of higher-twists (e.g., twist-5 and twist-6),
  sub-leading Fock states and higher helicity components whose contributions 
  should be tiny, not exceeding even 1\%.
\item  Our LO (in $\alpha_s$) scattering kernels are apparently gauge dependent 
arising from the contributions of the single hard gluon propagator. However, in 
the context of the $\pi\gamma^*\rightarrow\gamma$ 
transition form factor, it can be shown through a systematic order by order 
calculation using $k_T$-factorization that there is indeed a cancellation of the
gauge dependences between the quark-level diagrams of the hard kernel and the 
effective diagrams of the pion wavefunction \cite{Nandi}, so that the net
result turns out to be gauge-invariant to all orders. It is, thus, believed
that the same technique can be straightforwardly extended to other hadronic 
elastic and transition form factors, including the present context of the kaon 
form factors, at least up to the level of NLO corrections.
\item The factorized hard form factors further suffer from renormalization/factorization 
scale dependent ambiguities that typically emerge from the truncation of the
perturbative series and would be absent if we were able to obtain an all-order
result in the QCD coupling $\alpha_s$. To minimize the scale dependence in our
present investigation, we adhere to a fixed prescription with the scales set
to the momentum transfer $Q$ \cite{Coriano,Nandi}, as mentioned previously 
in the context of the Sudakov factor. In this way, we hope to improve the 
reliability and self-consistency of the perturbative prediction and reduce 
the influence from higher-order corrections. 
\item Nevertheless, a naive estimation of the NLO twist-2 contributions to 
the kaon form factors, using available NLO calculations for the pion form 
factor in asymptotic QCD, can be roughly expressed as \cite{NLO}   
\begin{equation} 
Q^2\{F,G\}^{\rm NLO}_K\approx(0.903\,{\rm GeV}^2)\,\alpha^2_s(Q^2)\,\frac{f^2_K}{f^2_\pi}\,, 
\end{equation}
which yields a rather nominal contribution $\sim 20-30\%$ that is roughly of
the same order of magnitude as the twist-4 contributions obtained in our
analysis. It is to be noted that the above estimation is based on 
the usual collinear factorization approach \cite{NLO} which did not take 
TMD into account. Including the $k_T$ dependence of the kernel might 
further reduce the NLO corrections, as was shown in the cases of the 
pion \cite{k_T_pi} and the nucleon \cite{k_T_N} form factors. It goes 
without saying that a full systematic NLO calculation (including twists-3) 
within the TMD factorization scheme, which is missing until now, 
would be indispensable in resolving this issue about the definitive 
magnitude of the sub-leading corrections.   
\end{itemize}

On the experimental side, as seen in Fig.~\ref{fig:total_FF}, currently 
the space-like region is completely devoid of data points at $Q^2$ values higher 
than $\sim0.2$ GeV$^2$. This makes it difficult, if not impossible, to compare 
such low-energy data with our predictions based on a pQCD approach which
becomes unreliable and diverges rapidly in the vicinity of the Landau pole 
$\Lambda_{\rm QCD}\approx0.2$ GeV. For the time-like region, there existed 
some older kaon data at relatively higher energies but with very poor
statistics \cite{ee-hh}. For such measurements the data above $Q^2>4.7$
GeV$^2$ had either upper limits or errors $\geq 50\%$. However, the 
recent CLEO measurements \cite{CLEO} at $Q^2=13.48$ GeV$^2$, apparently with a 
very small error-bar of $\pm 15\%$, can provide first possible opportunity to 
critically test theoretical predictions, although they do not shed light on
the variation with $Q^2$, which is a distinguishing feature of our result. 
Clearly, not only the moderate-energy time-like data seems reasonably reconciled, 
at higher energies both the CLEO result and the recent phenomenological 
prediction from $J/\psi$ decays: $M_{J/\psi }^{2}\left\vert G_{K
}(M_{J/\psi}^{2}\!=\!9.59\,\mathrm{GeV}^{2})\right\vert=0.81\pm0.18$
GeV$^2$ \cite{Seth}, lie within reasonable range of our prediction for the 
total time-like form factor. This is surprisingly consistent with the pion 
form factor results presented in \cite{Raha2}, also obtained within the 
light-cone $k_T$-factorization approach, that agreed well with most of 
the available moderate-energy data (with statistics far better 
than the kaon data), including the CLEO data and a similar phenomenological 
prediction \cite{Milana} based on $J/\psi$ decay analysis. Note that in  
the analysis \cite{Raha2}, the central value of the twist-3 chiral parameter 
$\mu_\pi$ was also taken to be 1.5 GeV.

\begin{figure}
    \hspace{-0cm}\begin{tabular}{cc}
      \resizebox{13.5cm}{!}{\includegraphics{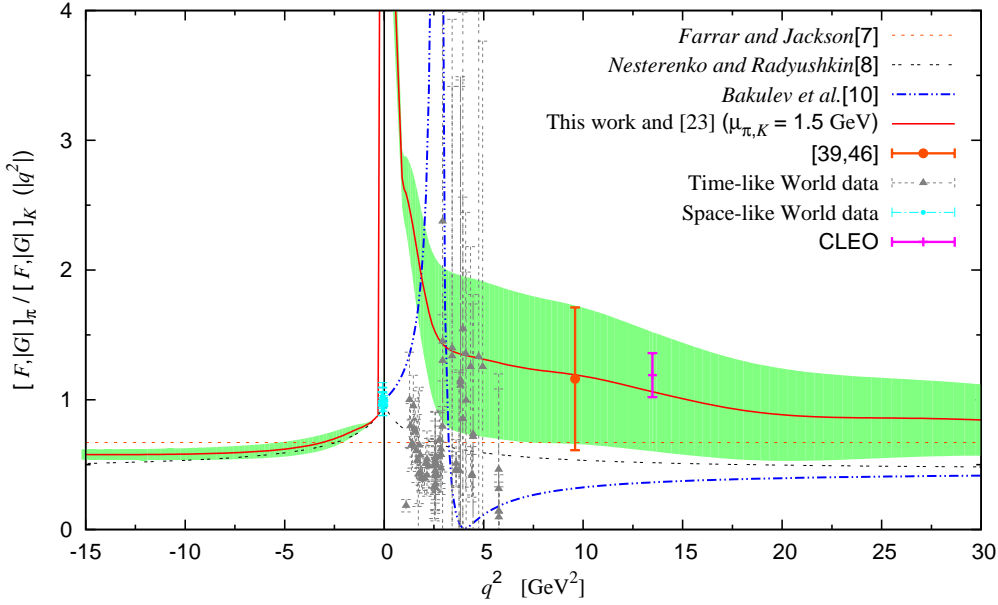}} 
    \end{tabular}
    \caption{Variation of the ratio of the pion and kaon time-like form
     factors with $Q^2$ in different approaches. The pion and kaon
     experimental data are taken from \cite{pion_exp,ee-hh,CLEO,kaon_exp_space}.}
    \label{fig:ratio_FF}
\end{figure}  

To this end, we consider the pion to kaon form factor ratios. In 
Fig.~\ref{fig:ratio_FF}, using the central result for the pion form factors 
from \cite{Raha2} (with $\mu_\pi=1.5$ GeV) we plot its variation with $Q^2$. 
The theoretical errors of the present work and \cite{Raha2} are added in 
quadrature to obtain the error band as shown in the figure. The large error 
should not come as a surprise as the errors of both the pion and kaon factors 
are large. We now compare this result with other theoretical predictions and 
available experimental data. Note that the standard asymptotic pQCD result of 
{\it Farrar and Jackson} \cite{FJ} yields a $Q^2$ independent ratio,
\begin{equation}
\left|\frac{\{F,G\}^{\rm asy}_{\pi}(Q^2)}{\{F,G\}^{\rm asy}_{K}(Q^2)}\right|=\frac{f^2_{\pi}}{f^2_{K}}=0.67\,.
\end{equation}
Clearly, the central value of our time-like ratio deviates appreciably 
from the asymptotic value at low- and moderate-$Q^2$, but however, it 
gradually approaches the asymptotic value at large $Q^2$, and so does the
space-like ratio. While our prediction fails to agree with the very low-energy 
time-like data points \cite{ee-hh}, showing the limitations of pQCD at such 
low-$Q^2$ values, the higher $Q^2\sim 4$ GeV$^2$ data points can somewhat 
be accommodated within our error-bars. At the same time, our time-like 
ratio at $Q^2=13.48$ GeV$^2$, i.e.,
\begin{equation}
\left|\frac{G_{\pi}(13.48 \,{\rm GeV}^2)}{G_{K}(13.48 \,{\rm
    GeV}^2)}\right|=1.06\pm0.46\,,
\end{equation}
is surprisingly close to the CLEO value: $1.19\pm0.17$ at $Q^2=13.48$ 
GeV$^2$ \cite{CLEO}, and the result obtained by taking the ratio of the
phenomenologically estimated time-like pion form factor \cite{Milana}
and the time-like kaon form factor \cite{Seth}, with the respective errors
again added in quadrature: 
\begin{equation}
\left|\frac{G_{\pi}(M^2_{J/\psi}=9.48\,{\rm GeV}^2)}{G_{K}(M^2_{J/\psi}=9.48\,{\rm
    GeV}^2)}\right|=1.16\pm0.55
\end{equation}
It is also noteworthy mentioning that the recent analysis \cite{DFM} based on a 
Light-Front Covariant Model (LFCM) up to $-q^2=10$ GeV$^2$, yielded the ratio
of the form factors quite similar to what we obtain in the space-like region. 
Finally, in Fig.~\ref{fig:ratio_FF}, we compare our result, evidently 
working better towards large-$Q^2$ values, with the soft QCD results
obtained from QCDSR which are instead known to yield reliable predictions 
at low- and moderate-$Q^2$ values. For example, the plot corresponding to the 
LD result \cite{Nest} not only agrees well with the very low-$Q^2$ space-like data 
(not resolved in the figure), but also with the low-energy time-like data 
when naively used in the time-like region. While, the analytically continued 
time-like LD result (see, Eq.~\ref{eq:soft_an}) \cite{Stefanis1} at
low-energies yields a plot very different from that of \cite{Nest}, but toward 
larger-$Q^2$ values both yield very similar predictions. Nevertheless, 
the QCDSR results significantly differ from the CLEO result and the 
one obtained from the phenomenological $J/\psi$ decay analysis. It is 
to be noted that in spite of the additional inclusion of the hard 
contributions, our space-like ratio of the total form factors does not 
differ significantly from that of \cite{Nest}, except at the very 
low-$Q^2\sim 0.2$ GeV$^2$ below which our result rapidly blows up. 

To sum up, in this paper we tried to systematically study the higher twist 
effects, namely, the twist-3 and twist-4 corrections to the standard twist-2 
pQCD charged kaon form factors by adopting minimal model dependence arising 
from the inclusion of (a) the transverse degrees of freedom in the kaon 
wavefunctions/DAs, and (b) the non-factorizable soft QCD corrections via 
local duality. The work presented here extends and completes the 
analyses of the previous work \cite{Raha1,Raha2}. Assuming the validity of the 
$k_T$-factorization ansatz through the explicit TMD of the scattering kernel, 
we showed a non-trivial twist-3 contribution in the 2-particle sector 
which along with the large soft QCD corrections turn out to be the real 
hallmark of the ``modified pQCD + soft QCD'' approach to determine the space-
and time-like kaon form factors. Other correction such as the 2-particle 
twist-4 were explicitly shows to have minor contributions only. To this end, 
the available moderate-energy experimental kaon data seems to be reasonably 
reconciled with the range of our predictions. It is also reassuring that 
the same approach works equally well independently for the electromagnetic 
pion form factors, which adds confidence to the arguments used in obtaining 
our results. It may, therefore, be speculated why the factorized result works 
so well for both the pion and kaon form factors in obtaining estimates, at
least in the correct ``ball-park'', in spite of factors like the resonances, 
hadronization and other final state interaction, naively neglected in this 
approach, that may render the factorized pQCD result questionable at the 
presently probed phenomenological region. However, to draw definite 
conclusion it is invaluable to have more high precision intermediate 
energy data, rather than to base our conclusions on such poor quality data.    
          
\noindent {\it Acknowledgments:} The authors would like to thank J.-W Chen and
H.-N. Li and N.G. Stefanis for various discussions. 

\pagebreak
 

\section{Appendix}

\setcounter{equation}{0}

\subsection*{2-particle Collinear Distribution Amplitudes (DAs)}
\label{app:DA}
The 2-particle twist-3 collinear DAs for the charged kaon (say, $K^-$) are defined 
at the scale of $\mu_0\approx 1$ GeV, in terms of  the following non-local matrix 
elements \cite{Filyanov,Ball,Lenz} :
\begin{eqnarray}
\langle0|\bar{u}(z)\,i\gamma_5\, s(-z)|K^-(P)\rangle\!\! &=&\!\!
\mu_K\int^1_0 dx\,e^{i\xi(Pz)}\phi^{\,p}_{3;K}(x,\mu_0^2)\,,\\
\langle0|\bar{u}(z)\,\sigma_{\alpha\beta}\gamma_5\,s(-z)|K^-(P)\rangle\!\!&=&\!\!-\frac{i}{3}\,\mu_K
\left\{1-\left(\frac{m_u+m_s}{m_K}\right)^2\right\}\nonumber\\
&&\times\,(P_\alpha z_\beta -P_\beta z_\alpha)\,\int^1_0 dx\,e^{i\xi(Pz)}\phi^\sigma_{3;K}(x,\mu_0^2)\nonumber
\end{eqnarray}
with $\mu_K=m^2_K/(m_u+m_s)$ and $\xi=2x-1$. Note that the gauge-link factors 
(Wilson-line) in the matrix elements are to be implicitly understood. The normalization 
conditions for the above twist-3 DAs are given by
\begin{equation}
\label{eq:norm-twist3}
N^{\,p,\sigma}_{3;K}=\int^1_0\,dx\,\phi^{\,p,\sigma}_{3;K}(x,\mu^2)=\frac{f_{K}}{2\sqrt{2N_c}}
\,,
\end{equation}
which have the following asymptotic forms:
\begin{eqnarray}
\label{eq:twist-3asy}
\phi^{{p}\,({\rm as})}_{3;K}(x)\!\!&=&\!\!\frac{f_{K}}{2\sqrt{2N_c}}\,,
\nonumber\\
\phi^{{\sigma}\,({\rm as})}_{3;K}(x)\!\!&=&\!\!\frac{3f_{K}}{\sqrt{2N_c}}\,x
(1-x)\,.
\end{eqnarray}
The explicit formulas for the non-asymptotic 2-particle twist-3 collinear 
DAs, expressed as a series expansion over conformal spins at 
next-to-leading order, are given by \cite{Ball}
\begin{eqnarray}
\phi^p_{3;K}(x,\mu^2)\!\!&=&\!\!\phi^{{p}\,({\rm as})}_{3;K}(x)\left\{1+\left(30\eta_{3K}(\mu^2)-\frac{5}{2}\,\rho^2_K(\mu^2)\right)C^{1/2}_2(\xi)\right.\nonumber\\
&&+\left.\left(-3\eta_{3K}(\mu^2)\,\omega_{3K}(\mu^2)-\frac{27}{20}\,\rho^2_K(\mu^2)-\frac{81}{10}\,\rho^2_K(\mu^2)\,a^{K}_2(\mu^2)\right)C^{1/2}_4(\xi)\right\}\,,\nonumber
\end{eqnarray}
\begin{eqnarray}
\phi^\sigma_{3;K}(x,\mu^2)\!&=&\!\phi^{{\sigma}\,({\rm as})}_{3;K}(x)\left\{1+\left(5\eta_{3K}(\mu^2)-\frac{1}{2}\eta_{3K}(\mu^2)\,\omega_{3K}(\mu^2)\right.\right.\nonumber \\
&&-\left.\left.\frac{7}{20}\,\rho^2_K(\mu^2)-\frac{3}{5}\,\rho^2_K(\mu^2)
   \,a^{K}_2(\mu^2)\right)C^{3/2}_2(\xi)\right\}
\end{eqnarray}
with
\begin{eqnarray}
\eta_{3K}=\frac{f_{3K}}{f_K}\frac{1}{\mu_K}\quad ;\quad
\rho_K=\frac{m_K}{\mu_K}\,,\nonumber
\end{eqnarray}
the non-perturbative parameters $f_{3K}$ and $\omega_{3K}$ being
defined through the following matrix elements of local twist-3
operators:
\begin{eqnarray}
\langle0|{\bar u}(0)\,\sigma_{\mu\nu}\gamma_5\, g_sG_{\alpha\beta}\,s(0)|
K^-(P)\rangle=if_{3K}\left( P_\alpha P_\mu g_{\nu\beta}-
P_\alpha P_\nu g_{\mu\beta} - P_\beta P_\mu g_{\nu\alpha}+ P_\beta P_\nu
g_{\alpha\mu}\right) \,, \nonumber
\end{eqnarray}
\begin{eqnarray}
\langle 0|{\bar
  u}(0)\,\sigma_{\mu\lambda}\gamma_5[iD_\beta,g_sG_{\alpha\lambda}]\,s(0)\!\!&-&\!\!\frac{3}{7}\,
i\partial_\beta\,{\bar u}(0)\,\sigma_{\mu\lambda}\gamma_5\,g_sG_{\alpha\lambda}\,s(0)|K^-(P)\rangle\nonumber \\
&&\hspace{-2cm}=\frac{3}{14}\,if_{3K}P_\alpha P_\beta P_\mu\,\omega_{3K}+{\mathcal O}(\rm higher\,\,twist)\,,
\end{eqnarray}
where $g_s$ is the strong coupling and $G_{\alpha\beta}$ is the gluon field tensor. 
The LO scale dependence of various twist-3 parameters are given by
\begin{eqnarray}
\label{eq:fw_3K}
\rho_{K}(\mu^2)\!&=&\!L^{\gamma^{(0)}_{3;s{\bar u}}/\beta_0}\,\rho_{K}(\mu^2_0)\,;
\quad\gamma^{(0)}_{3;s{\bar u}}=1\,, \nonumber\\
\eta_{3K}(\mu^2)\!&=&\!L^{\gamma^{(0)}_{3;\eta}/\beta_0}\,\eta_{3K}(\mu^2_0)\,;\quad
\gamma^{(0)}_{3;\eta}=\frac{4}{3}\,{\mathcal C}_F + \frac{1}{4}\,{\mathcal C}_A
\,,\nonumber \\
\omega_{3K}(\mu^2)\!&=&\!L^{\gamma^{(0)}_{3;\omega}/\beta_0}\,\omega_{3K}(\mu^2_0)\,;\quad
\gamma^{(0)}_{3;\omega}=-\frac{7}{24}\,{\mathcal C}_F + \frac{7}{12}\,{\mathcal C}_A\,,\nonumber\\
a^{K}_1(\mu^2)\!&=&\!L^{\gamma^{(0)}_1/\beta_0}\,a^{K}_1(\mu^2_0)\,;\quad
\gamma^{(0)}_1= \frac{2}{3}\,{\mathcal C}_F\,,\nonumber \\
a^{K}_2(\mu^2)\!&=&\!L^{\gamma^{(0)}_2/\beta_0}\,a^{K}_2(\mu^2_0)\,;\quad
\gamma^{(0)}_2= \frac{25}{24}\,{\mathcal C}_F\,,
\end{eqnarray}
where $L=\alpha_s(\mu^2)/\alpha_s(\mu^2_0)$,\, ${\mathcal
  C}_F=(N^2_c-1)/2N_c$ and ${\mathcal C}_A=N_c$.
However, the strange quark being massive, there is operator mixing of the ones in
Eq.~(\ref{eq:fw_3K}) with those of twist-2 operators, so that the resulting LO
RG equations give the following scale dependences:
\begin{eqnarray}
f_{3K}(\mu^2)\!&=&\!L^{55/36\beta_0}f_{3K}(\mu^2_0)+\frac{2}{19}\left(L^{1/\beta_0}-L^{55/36\beta_0}\right)[f_K\,m_s](\mu^2_0)
\nonumber\\
&&+\,\frac{6}{65}\left(L^{55/36\beta_0}-L^{17/9\beta_0}\right)[f_K\,m_s\,a^K_1](\mu^2_0)\,,\nonumber\\
\left[f_{3K}\,\omega_{3K}\right](\mu^2)\!&=&\!L^{26/9\beta_0}[f_{3K}\,\omega_{3K}](\mu^2_0)
+\frac{1}{170}\left(L^{1/\beta_0}-L^{26/9\beta_0}\right)[f_K\,m_s](\mu^2_0)\nonumber\\
&&+\,\frac{1}{10}\left(L^{17/9\beta_0}-L^{26/9\beta_0}\right)[f_K\,m_s\,a^K_1](\mu^2_0)\nonumber\\
&&+\,\frac{2}{15}\left(L^{43/(18\beta_0)}-L^{26/9\beta_0}\right)[f_K\,m_s\,a^K_2](\mu^2_0)\,.
\end{eqnarray}

The 2-particle twist-4 collinear DAs modify the twist-2 axial matrix 
element and are given by
\begin{eqnarray}
\langle0|\bar{u}(z)\,\gamma_\mu
\gamma_5\,s(-z)|K^-(P)\rangle\!&=&\!iP_\mu\,\int^1_0 dx\,e^{i\xi(P z)}
\,\left[\phi_{2;K}(x,\mu_0^2)+\frac{1}{4}m^2_K z^2 {\mathbb A}_{4;K}(x,\mu^2_0)\right]\nonumber\\
&&+\frac{i}{2}f_K m^2_K\frac{1}{Pz} z_\mu\,\int^1_0 dx\,e^{i\xi(P z)}\,{\mathbb B}_{4;K}(x,\mu_0^2)\,,
\end{eqnarray}
where  ${\mathbb{B}}_{4;K}=g_{4;K}-\phi_{2;K}$, with the normalization
conditions expressed as
\begin{eqnarray}
\label{eq:norm-twist4}
N^{\,{\mathbb{A}},g}_{4;K}=\int^1_0\,dx\,\{{\mathbb A},g\}_{4;K}(x,\mu^2)=\frac{f_{K}}{2\sqrt{2N_c}}\,,
\end{eqnarray}
and the asymptotic forms namely,
\begin{eqnarray}
\label{eq:twist-4asy}
{\mathbb A}^{({\rm as})}_{4;K}(x)\!&=&\!\frac{15f_{K}}{\sqrt{2N_c}}\,x^2(1-x)^2\,,
\nonumber\\
g^{({\rm as})}_{4;K}(x)\!&=&\!\frac{f_{K}}{2\sqrt{2N_c}}\,.
\end{eqnarray}
Next we display the explicit forms of the non-asymptotic twist-4 collinear 
DAs at NLO in conformal spin \cite{Ball}:
\begin{eqnarray}
{\mathbb A}_{4;K}(x,\mu^2)\!&=&\!\frac{3f_{K}}{{\mathcal
    N}_{\mathbb{A}}\sqrt{2N_c}}\,x(1-x)\left\{\frac{16}{15}+\frac{24}{35}
\,a^{K}_2(\mu^2)+20\,\eta_{3K}(\mu^2)+\frac{20}{9}\,\eta_{4K}(\mu^2)\right.\nonumber\\
&&\hspace{-1.5cm}\left.+\left(-\frac{1}{15}+\frac{1}{16}-\frac{7}{27}\,\eta_{3K}(\mu^2)\,\omega_{3K}(\mu^2)-\frac{10}{27}\,
\eta_{4K}(\mu^2)\right)\,C^{3/2}_2(\xi)+\left(-\frac{11}{210}\,a^{K}_2(\mu^2)\right.\right.\nonumber\\
&&\hspace{-1.6cm}\left.\left.-\frac{4}{135}\,\eta_{3K}(\mu^2)\,\omega_{3K}(\mu^2)\right)\,C^{3/2}_{4}(\xi)\right\}
+\frac{f_{K}}{2{\mathcal
    N}_{\mathbb{A}}\sqrt{2N_c}}\,\left(-\frac{18}{5}\,a^{K}_{2}(\mu^2)+21\eta_{4K}(\mu^2)\,
\omega_{4K}(\mu^2)\right)\nonumber\\
&&\hspace{-1.5cm}\times\left\{2x^3(10-15x+6x^2)\ln{x}+2\bar{x}^3(10-15\bar{x}+6\bar{x}^2)\ln\bar{x}+
x\bar{x}(2+13x\bar{x})\right\}\,,\nonumber
\end{eqnarray}
\begin{eqnarray}
g_{4;K}(x,\mu^2)\!&=&\!\frac{f_{K}}{2\sqrt{2N_c}}\left\{1+\left(1+\frac{18}{7}\,a^{K}_2(\mu^2)+60\,\eta_{3K}(\mu^2)
+\frac{20}{3}\,\eta_{4K}(\mu^2)\right)\,C^{1/2}_2(\xi)\right.\nonumber\\
&&\left.+\left(-\frac{9}{28}\,a^{K}_2(\mu^2)-6\,\eta_{3K}(\mu^2)\,\omega_{3K}(\mu^2)\right)\,C^{1/2}_4(\xi)\right\}\,,
\end{eqnarray}
where $\bar{x}=1-x$, and in the notation of \cite{Filyanov}, 
$\delta^2\equiv m^2_K\eta_{4K}$ and $\epsilon\equiv21/8\,\omega_{4K}$. 
Note that the additional factor ${\mathcal N}_{\mathbb{A}}\approx 3.5$ in the denominator of ${\mathbb A}_{4;K}$ 
is a contrast to the expression given in \cite{Ball}, which is introduced to 
normalize the DA. The non-perturbative parameters $\eta_{4K}$ and $\omega_{4K}$ are 
defined through the following matrix elements of local twist-4 operators:
\begin{eqnarray}
\langle0|{\bar u}(0)\,\gamma_\alpha\, ig_s\widetilde{G}_{\mu\nu}\,s(0)
|K^-(P)\rangle=-\frac{1}{3}f_{K}m^2_K\eta_{4K}\left( P_\mu g_{\nu\alpha}-
P_\nu g_{\mu\alpha}\right) \,, \nonumber
\end{eqnarray}
\begin{eqnarray}
\langle0|{\bar
  u}(0)\,[iD_\mu,ig_s\widetilde{G}_{\nu\lambda}]\gamma_\lambda\,s(0)\!\!&-&\!\!\frac{4}{9}\,i\partial_\mu\,{\bar
  u}
(0)\,ig_s\widetilde{G}_{\nu\lambda}\gamma_\lambda\,s(0)|K^-(P)\rangle\nonumber \\
&&\hspace{-3.3cm}=\!f_{K}m^2_{K}\eta_{4K}\omega_{4K}\left(P_\mu P_\nu -
\frac{1}{4}m^2_{K}\,g_{\mu\nu}\right)+{\mathcal O}
(\rm{twist \,5})\,,
\end{eqnarray}
where, $\widetilde{G}_{\mu\nu}=\frac{1}{2}\epsilon_{\mu\nu\rho\sigma}G^{\rho\sigma}$ is the dual gluon field tensor.
Taking into account the mixing with the operators of lower twists, the  LO 
RG evolution of the twist-4 parameters are
\begin{eqnarray}
\eta_{4K}(\mu^2)\!&=&\!L^{\gamma^{(0)}_{4;\eta}/\beta_0}\eta_{4K}(\mu^2_0)
+\frac{1}{8}\left(1-L^{\gamma^{(0)}_{4;\eta}/\beta_0}\right)\,;\quad\gamma^{(0)}_{4;\eta}=\frac{2}{3}\,{\mathcal C}_F\,, \nonumber\\
\left[\eta_{4K}\,\omega_{4K}\right](\mu^2)\!&=&\!L^{\gamma^{(0)}_{4;\eta\omega}/\beta_0}\left[\eta_{4K}\,
\omega_{4K}\right](\mu^2_0)\,;\quad\gamma^{(0)}_{4;\eta\omega}=\frac{5}{6}\,{\mathcal C}_A\,.
\end{eqnarray}
Finally, we present the various Gegenbauer polynomials used in the above formulas:
\begin{eqnarray}
C^{1/2}_0(\xi)\!&=&\!1\,,\nonumber\\
C^{1/2}_1(\xi)\!&=&\!\xi\,,\nonumber\\
C^{1/2}_2(\xi)\!&=&\!\frac{1}{2}\,(3\xi^2-1)\,,\nonumber\\
C^{1/2}_3(\xi)\!&=&\!\frac{1}{2}\,\xi(5\xi^2-3)\,,\nonumber\\
C^{1/2}_4(\xi)\!&=&\!\frac{1}{8}\,(35\xi^4-30\xi^2+3)\,,\nonumber\\
C^{3/2}_0(\xi)\!&=&\!1\,,\nonumber\\
C^{3/2}_1(\xi)\!&=&\!3\xi\,,\nonumber\\
C^{3/2}_2(\xi)\!&=&\!\frac{3}{2}\,(5\xi^2-1)\,,\nonumber\\
C^{3/2}_3(\xi)\!&=&\!\frac{5}{2}\,\xi(7\xi^2-3)\,.
\end{eqnarray}



\end{document}